\documentclass[preprint,review,12pt,number]{elsarticle}

\usepackage[english]{babel}
\usepackage{color, graphicx}
\usepackage{soul}
\usepackage{caption, subcaption}
\usepackage{amsmath}
\usepackage{xcolor}
\usepackage{booktabs}
\graphicspath{{figures/}}
\usepackage{epstopdf}
\usepackage{dsfont}
\usepackage[utf8]{inputenc}
\usepackage [autostyle, english = american]{csquotes}
\MakeOuterQuote{"}

\usepackage{dblfloatfix}
\usepackage[top=2.5cm,bottom=2.5cm,left=2.5cm,right=2.5cm]{geometry}
\journal{Journal of Wind Engineering and Industrial Aerodynamics}

\begin{document}
\def\floatpagepagefraction{1}
\def\textpagefraction{.001}

% Main title of the paper
\title {Uncertainty Quantification of a Wind Tunnel-Informed Stochastic Wind Load Model for Wind Engineering Applications}          % 

\author[UF]{Thays Guerra Araujo Duarte}
\ead{tduarte.guerraar@ufl.edu}
\author[Umich]{Srinivasan Arunachalam}
\ead{sriarun@umich.edu}
\author[UF]{Arthriya Subgranon\corref{cor1}\fnref{fn1}}
\ead{Arthriya.Subgranon@essie.ufl.edu}
\author[Umich]{Seymour M.J. Spence}
\ead{smjs@umich.edu}
\cortext[cor1]{Corresponding author}
\fntext[fn1]{Tel. +1-352-392-9537, Fax +1-352-392-3394}

\address[UF]{Department of Civil and Coastal Engineering, University of Florida, Gainesville, FL 32611, USA}
\address[Umich]{Department of Civil and Environmental Engineering, University of Michigan, Ann Arbor, MI 48109, USA}

\begin{abstract}

The simulation of stochastic wind loads is necessary for many applications in wind engineering. The proper orthogonal decomposition (POD)-based spectral representation method is a popular approach used for this purpose due to its computational efficiency. For general wind directions and building configurations, the data-driven POD-based stochastic model is an alternative that uses wind tunnel smoothed auto- and cross-spectral density as input to calibrate the eigenvalues and eigenvectors of the target load process. Even though this method is straightforward and presents advantages compared to using empirical target auto- and cross-spectral density, the limitations and errors associated with this model have not been investigated. To this end, an extensive experimental study on a rectangular building model considering multiple wind directions and configurations was conducted to allow the quantification of uncertainty related to the use of wind tunnel data for calibration and validation of the data-driven POD-based stochastic model. Errors associated with the use of typical wind tunnel records for model calibration, the model itself, and the truncation of modes were quantified. Results demonstrate that the data-driven model can efficiently simulate stochastic wind loads with negligible model errors, while the errors associated with calibration to typical wind tunnel data can be important.
\end{abstract}


% Use if graphical abstract is present
% \begin{graphicalabstract}
% \includegraphics{figs/grabs.pdf}
% \end{graphicalabstract}
% Keywords
% Each keyword is seperated by \sep

\maketitle

\section{Introduction}

Over the last decade, several studies have focused on advancing performance-based wind engineering (PBWE), with multiple frameworks being proposed (e.g., \citep{ciampoli2011performance,smith2011monte,petrini2012performance,barbato2013performance,spence2014performance,bernardini2015performance,chuang2017performance,cui2018unified,ouyang2020performance,Cui_2020,Ouyang_21,Arunachalam_2022,chuang2022framework}), which enable the incorporation and propagation of uncertainties during the risk assessment of wind-excited systems. In this context, wind loads are among the most important sources of uncertainty. Wind tunnel tests are commonly used to characterize the effects of wind on structures, providing detailed information on complex building-specific aerodynamic phenomena \citep{lin2005characteristics, kareem2013advanced}. However, multiple measurements, and/or long-duration experimental tests that aim at characterizing the uncertainties involved in the process are usually limited due to costs. Given these limitations, the use of numerically simulated processes is a powerful alternative to reduce the reliance on experimental data for the analysis of a variety of engineering problems. 

Simulation algorithms for the generation of stationary multivariate stochastic processes have been proposed, which are based on the spectral representation method (SRM) and have the advantage of being computationally efficient (e.g.,  \citep{shinozuka1971simulation,shinozuka1991simulation,deodatis1996simulation}). Among many simulation methods, the proper orthogonal decomposition (POD)-based SRM has become popular, especially in the wind engineering field, with many studies showing its computational efficiency as well as its ability to extract dominant patterns in random fluctuations of the measured wind field \citep{tamura1999proper,carassale2001double,chen2005simulation,chen2005proper,huang2020data,ouyang2020performance}. Another advantage of the POD-based stochastic wind load model compared to the direct SRM is the possibility of truncation of higher-order modes. However, while truncation accelerates the simulation, it can influence the accuracy of the simulated signals if an insufficient number of modes are considered  \citep{chen2005proper,tao2020error}. Therefore, truncation errors are introduced within the simulation, affecting the probabilistic characteristics of the generated wind fields/loads and, as a consequence, the structural response.

Previous studies have focused on quantifying the statistical errors, i.e., bias and stochastic errors, associated with the stochastic simulation of wind fields through the conditional-simulation method \citep{hu2010error}, and interpolation errors introduced by interpolation-enhanced SRM \citep{tao2018reduced, tao2020error}. 
The errors produced by the POD-based SRM are estimated in \citep{hu2010error} and compared to the Cholesky decomposition-based errors, where observations suggest the stochastic errors are smaller for a POD-based decomposition as compared to a Cholesky-based decomposition.

The realizations of a stochastic model need to match the probabilistic characteristics (e.g., second-order statistics) of a target process. Several discussions on analytical formulations of the power spectral density (PSD) and cross power spectral density (CPSD) have been published, in which empirical relations based on a large amount of data corresponding mostly to prismatic shape structures, and alongwind and crosswind dynamic responses have been proposed \citep{davenport1971response,simiu1974wind,melbourne1980comparison,solari1983analytical, kaimal1994atmospheric}. However, due to the complexity involved in establishing mathematical models that characterize analytical PSD/CPSD functions for general wind loading, approximations based on various assumptions are often considered. While previous studies have mostly used such analytical PSD/CPSD functions to calibrate the stochastic models \citep{chen2005proper, kareem2008numerical, kareem2013advanced, tao2020error}, some attention has been drawn to also using measured data as input (sample-based) to calibrate the stochastic process \citep{gurley1998simulation, wang2013data,suksuwan2018optimization,huang2020data,ouyang2020performance}. Using wind tunnel data spectral functions as a prescribed target in simulation models is advantageous when compared to using analytical PSD/CPSD models due to the fact that the main aerodynamic features can be captured for different building geometries, approaching flows, and configurations. As an example, wind tunnel testing is indispensable for the response characterization of dynamically sensitive tall buildings for which complex aerodynamic phenomena will generally occur (e.g., vortex shedding). The data-driven approach is advantageous as it ensures the complex phenomena captured in the wind tunnel are also captured in the stochastic simulation model.

A recent data-driven POD-based stochastic wind load model has been proposed as an alternative strategy to calibrating stochastic wind load models to analytical PSD/CPSD functions \citep{suksuwan2018optimization, ouyang2020performance}. Through the proposed approach, wind tunnel data is used to inform the spectral modes and eigenvalues of the target loading process. This method provides advantages such as its potential applicability for any wind direction and building geometry. In addition to that, this data-driven method aligns better with the recommendations of the ``Prestandard for Performance-based Wind Design" \citep{american2019prestandard}, which requires that wind tunnel testing be used to estimate wind loads for the design of engineered buildings. At the current state of knowledge, typically 30-second to 1-minute wind tunnel records are used for providing at least 1-hour of wind loading at full scale and confirm the assumption of stationarity of the wind loading \citep{american2019prestandard}. Previous studies have used this typical record length to calibrate stochastic wind load models. 

Regardless of the various advantages of wind tunnel-informed models, concerns still exist on the use of such models, and any possible limitations they may impose, due to a lack of experimental validation. For the experimental wind tunnel data, epistemic uncertainties, i.e., imperfect knowledge (e.g., model defects, human errors, low equipment resolution, sampling errors), and aleatory uncertainties, i.e., inherent randomness (e.g., variability of wind tunnel records collected for the same experimental setup), require characterization. As a consequence of these errors, the second-order statistics of the target process may not be fully captured. Although several experimental campaigns have been conducted to study the effects of different conditions on wind loading (e.g., different terrain roughness, surrounding structures, etc.), the errors associated with the estimation of the second-order properties of target stochastic processes using typical wind tunnel data have not been investigated thoroughly. If not accounted for, these errors can propagate through the stochastic wind load model, therefore, compromising a proper assessment of structural responses. 

In the present study, an extensive wind tunnel experimental campaign was conducted on a rectangular building model with three objectives: (1) quantify the errors associated with the variability of wind tunnel records in estimating PSD/CPSD functions; (2) assess how these errors affect the calibration of data-driven POD-stochastic wind models; and (3) quantify the model errors associated with data-driven POD-stochastic wind models. Multiple wind directions and two different experimental setups were considered. Errors associated with using a typical wind tunnel record (i.e., 32 seconds of recorded data) are quantified, as are the effects of wind direction and interference from surrounding buildings. Thus, this study was divided into four main parts: first, wind tunnel testing was carried out in order to identify the data-driven target PSD/CPSD functions of the wind loading process. Then, the errors associated with the variability induced through the use of a single typical wind tunnel record were investigated and compared to the target. The data-driven target PSD/CPSD functions are then used to calibrate the POD-based SRM and simulate realizations to investigate the propagation of the calibration errors to the simulation model. Finally, model errors are investigated as are errors induced by mode truncation. 

%--------------------------------------------------------------
%--------------------------------------------------------------

\section{Stochastic Wind Load Process}

A stochastic process is any random phenomenon varying over time that cannot be described by a closed-form equation, as each realization of the phenomenon will be unique. The wind loads are spatio-temporally varying random processes that can be modeled as stationary in time for most applications. Stationary wind loads have time-invariant probabilistic properties such as mean and covariance. In this study, it is assumed that wind loads can be adequately described by a stationary stochastic process. 

\begin{figure}
	\centering 
		\includegraphics[scale=0.45]{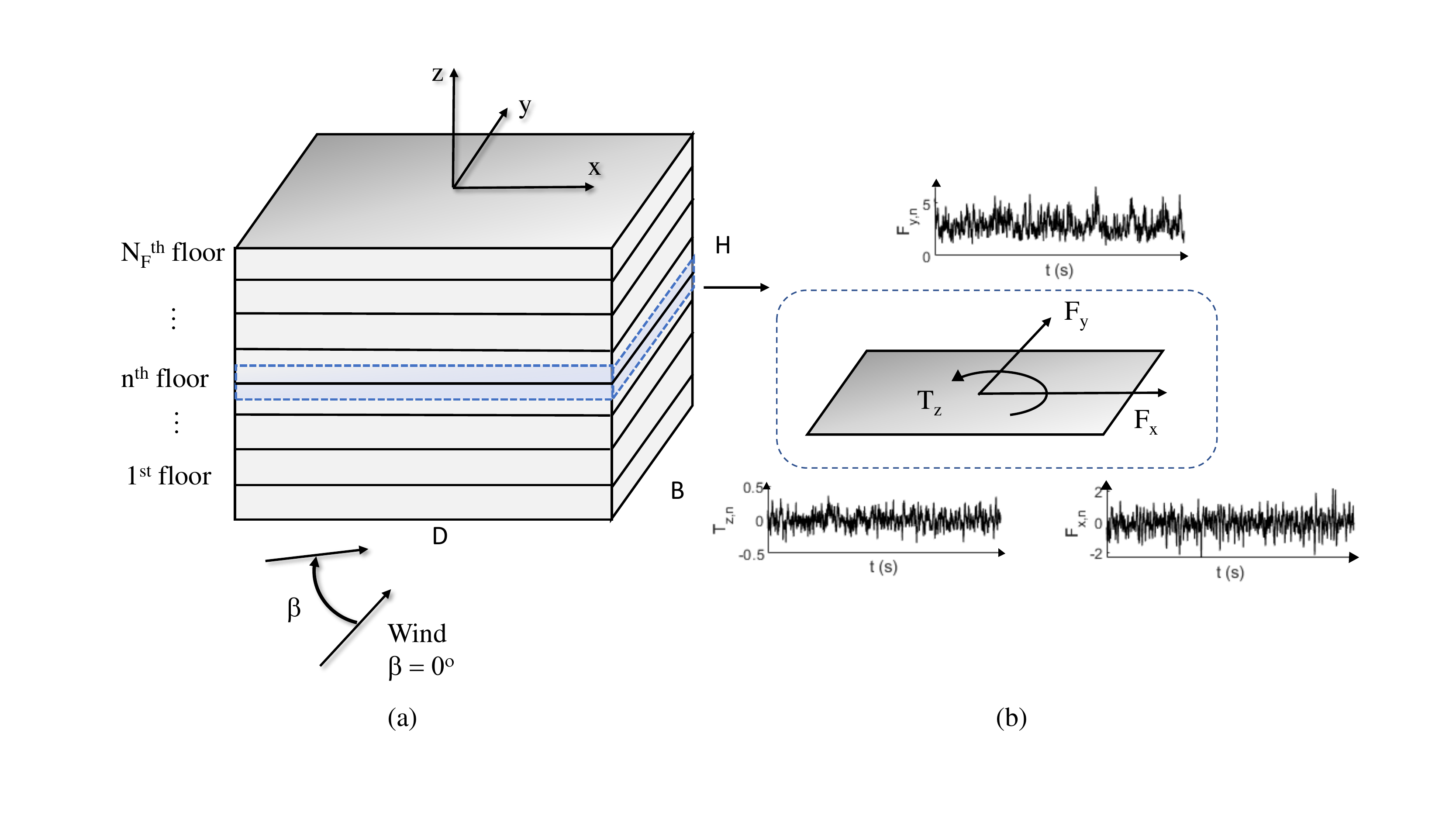}
	\caption{(a) Schematic representation of wind flow and tributary area of each floor, and (b) Resultant dynamic wind force and moment components.}
	\label{windcomponents}
\end{figure}

The wind load process can be divided into a mean and fluctuating (or turbulent) part. To model the fluctuating part, a zero-mean vector-valued wind load process can be defined: $\mathbf{P}(t)=[P_1(t),$  $P_2(t), ..., P_N(t)]^T$, where $N$ is the total number of components (e.g., generalized forces $F_x(t)$, $F_y(t)$, and $T_z(t)$ at the centroid of each floor of the building). Figures \ref{windcomponents} (a)-(b) illustrate the schematic representation of the generalized forces and wind directions considered in this study. The resultant dynamic wind loads at the $n$th story are obtained by integrating the pressures acting on appropriate tributary areas, where $F_{x,n}(t)$ and $F_{y,n}(t)$ represent the forces along $x$ and $y$ directions, and $T_{z,n}(t)$ represents the torsional moment around the $z$-axis at the centroid of the floor, as shown in Fig. \ref{windcomponents} (b). While wind pressures are known to have non-Gaussian features, the floor loads may be modeled as a Gaussian process due to the Central Limit Theorem \citep{gurley1997modelling}. Therefore, the knowledge of the variance and covariance are adequate for describing the stochastic wind load process, $\mathbf{P}(t)$, and will later be used to define the error measures.

%--------------------------------------------------------------
%--------------------------------------------------------------

\subsection{Proper Orthogonal Decomposition of Wind Process}
Proper orthogonal decomposition (POD) can be used to identify the dominant modes of a stochastic process, as well as increase the computational efficiency of SRM when simulating the process through the fast Fourier Transform (FFT) approach \citep{tamura1999proper,chen2005proper}.  

This technique allows for an efficient representation of the original random process, that can be fully coherent, through the decomposition into a set of non-coherent subprocesses. The decomposition can be carried out based on either the covariance or CPSD matrices using either Cholesky decomposition or Schur decomposition \citep{chen2005proper}. The latter is usually preferred since it allows a reduced number of modes to simulate the process, and is adopted in this study. 

A discrete zero-mean multivariate stationary random process for a given wind direction, $\beta$, can be mathematically expressed as $\mathbf{P}(t;\beta)=[P_1(t;\beta),$  $P_2(t;\beta), ..., P_N(t;\beta)]^T$, where $N$ is the total number of subprocesses. Through FFT, the time series is converted into the frequency domain, $\mathbf{P}(\omega;\beta)$, in which it can be decomposed in terms of normalized orthogonal mode functions, $\mathbf{\Psi}(\omega;\beta)$, as follows:
\begin{equation}
    \mathbf{P}(\omega;\beta)=\mathbf{\Psi}(\omega;\beta)\mathbf{a}(\omega;\beta) \\
    =\sum^{N}_{i=1}\mathbf{\Psi}_i(\omega;\beta)a_i(\omega;\beta)  
    = \sum^{N}_{i=1}\mathbf{P}_i(\omega;\beta)
\end{equation}
where $\mathbf{a}(\omega;\beta) $ is the fourier transform of the expansion series $\mathbf{a}(t;\beta)=[a_1(t;\beta),a_2(t;\beta), ..., a_N(t;\beta)]^T$; and $\mathbf{\Psi}(\omega;\beta)=[\mathbf{\Psi}_1(\omega;\beta),\mathbf{\Psi}_2(\omega;\beta),...,\mathbf{\Psi}_N(\omega;\beta)]$. The eigenvalues and eigenvectors of the multivariate stochastic process are obtained by solving the following eigenproblem:
\begin{equation} \label{eigenvalue}
    \left[\mathbf{S}_{\mathbf{P}}(\omega;\beta)-{\Lambda_i}(\omega;\beta)\mathbf{I}\right]\mathbf{\Psi_i}(\omega;\beta) =0
\end{equation}
where $\mathbf{S}_{\mathbf{P}}$ is the two-sided CPSD matrix of the zero-mean process $\mathbf{P}(t;\beta)$, which is Hermitian and non-negative definite \citep{shinozuka1987stochastic}; $\mathbf{I}$ is the identity matrix; and $\Lambda_i$ and $\mathbf{\Psi}_i$ are the $i$th frequency-dependent eigenvalue and eigenvector of $\mathbf{P}(t;\beta)$, respectively. Once the eigenvalue and eigenvectors are obtained, the CPSD matrix of $\mathbf{P}(t;\beta)$ can be expressed by the summation of the contributing modes as: 
\begin{equation}
    \mathbf{S_P}(\omega;\beta) = \sum^{N}_{i=1}\Lambda_i(\omega;\beta)\mathbf{\Psi}_i(\omega;\beta)\mathbf{\Psi}_i^*(\omega;\beta)
\end{equation}

One of the advantages of a POD representation of a stochastic process is the possibility of mode truncation. For many problems in wind engineering, the first few eigenmodes associated with the highest eigenvalues carry the majority of the energy of the process. The truncation of the higher modes enables the following reduced-order modeling of the process:
\begin{equation} \label{GP_Process}
    \hat{\mathbf{P}}(t;\beta)=\sum_{i=1}^{N_m}\mathbf{P}_i(t;\beta)
\end{equation}
where $N_m$ is the number of contributing modes such that $N_m\leq N$ \citep{chen2005proper}; and $\mathbf{P}_i(t;\beta)$ and $\mathbf{P}_i(\omega;\beta)$ are FFT pairs. A more detailed discussion on the CPSD-based POD can be found in \cite{chen2005proper}.

%--------------------------------------------------------------

\section{Data-Driven POD-Based Stochastic Wind Load Model}
\label{Sect_DataDrivenPOD}

Through using the CPSD matrix-based POD spectral representation method, each subprocess, $\tilde{\mathbf{{P}}}_i(t;\beta)$, can be simulated independently to obtain a multicorrelated Gaussian random process, ${\mathbf{P}}^{GP}(t;\beta)$ \citep{chen2005proper}. As an alternative to using empirical spectral models for calibrating the POD-based stochastic wind load model, \cite{suksuwan2018optimization} and \cite{ouyang2020performance} propose to calibrate the model using the second-order statistics of the wind load process, $\mathbf{P}(t;\beta)$, obtained directly from the wind tunnel. The method consists of generating ergodic samples of a stationary stochastic vector process, calibrated to a prescribed data-driven input CPSD matrix derived from a single typical wind tunnel record \citep{deodatis1996simulation}. 
 
This method has the unique advantage of incorporating building-specific aerodynamic phenomena (e.g., vortex shedding and separated flow) to the extent that they are captured in boundary layer wind tunnel data. The components acting in $x$, $y$, and $z$-rotational directions can, therefore, be simultaneously simulated through the use of this method. The model can be further enhanced through standardization of the input wind tunnel forces, or force coefficient time-series (see Appendix B), prior to applying the spectral POD method. A standardization scheme allows a rescaling of each data set to have zero mean and constant standard deviation, leading to a more even redistribution of energy within the POD, hence a lower number of modes is needed to accurately represent the component time series.

The stationary zero-mean stochastic subprocess, $\tilde{\mathbf{P}}_i(t;\beta)$, can be generated by:
\begin{equation} \label{Eq_subprocess}
    \mathbf{\tilde{P}}_i(t;\beta)=\sum_{k=0}^{N_l-1}{2}|\mathbf{\Psi}_i(\omega_k;\beta)|\sqrt{\Lambda_i(\omega_k;\beta)\Delta\omega}\cos(\omega_k t +\vartheta_k(\omega_k)+\theta_{ik}) 
\end{equation} 
where $\mathbf{\Psi}_i(\omega_k;\beta)$ and $\Lambda_i(\omega_k;\beta)$ are obtained as in Eq.(\ref{eigenvalue}) where $\mathbf{S_P}$ is obtained directly from a wind tunnel record, $N_l$ is the total number of discrete frequencies up to the Nyquist cutoff frequency, $\Delta\omega$ is the frequency increment with $\omega_k = k\Delta\omega$, $\theta_{ik}$ is a uniformly distributed independent random variable over [0,2$\pi$] characterizing the stochasticity of the process; and $\mathbf{\vartheta}_k(\omega_k)$ is the phase angle defined as:
\begin{equation}
    \mathbf{\vartheta}_k(\omega_k) = \tan^{-1}\left(\frac{\textrm{Im}(\mathbf{\Psi}_i(\omega_k;\beta))}{\textrm{Re}(\mathbf{\Psi}_i(\omega_k;\beta))}\right)
\end{equation}

The truncation of $\mathbf{P}^{GP}(t;\beta)$ to $N_m$ can be expressed as:
\begin{equation}
    \mathbf{P}^{GP}(t;\beta) \approx \mathbf{\hat{P}}^{GP}(t;\beta) = \mathbf{\overline{P}(\beta)} + \sum^{N_m}_{i=1}\mathbf{\tilde{P}}_i(t;\beta)
\end{equation}
where $\mathbf{\hat{P}}^{GP}(t;\beta)$ is the truncated representation of  $\mathbf{P}^{GP}(t;\beta)$, $\mathbf{\overline{P}}(\beta)$ is the mean wind load estimated from the wind tunnel record for a given wind direction $\beta$, and $\mathbf{\tilde{P}}_i(t;\beta)$ is the zero-mean subprocess of Eq. (\ref{Eq_subprocess}).

%-----------------------------------

\subsection{Establishment of the Data-Driven Input PSD/CPSD Functions} \label{target_spectra}

Prior to the simulation of the wind process, an essential step in preparing the data-driven POD-based stochastic wind model is obtaining the input PSD/CPSD functions to be used for the model calibration. The underlying, or target, PSD/CPSD functions would be an ideal input as it accurately represents the stochastic process. However, the lack of knowledge of building-specific true PSD/CPSD functions is often a significant challenge when using stochastic wind load models, especially when it comes to multiple wind directions and complex building geometries. Therefore, identifying representative data-driven input PSD/CPSD functions is necessary in order to enable the simulation of stochastic wind loads. This section outlines the method utilized to define the input PSD/CPSD functions from a single typical wind tunnel record for subsequent calibration of the eigenvalues and eigenmodes of the POD-based stochastic wind load model. Then, a discussion is reported on how the underlying, or target, PSD/CPSD functions can be obtained in an ideal case where multiple wind records are available.

%-------------------------------------

\subsubsection{Smoothed PSD/CPSD Functions} 
\label{spectra_welch}
For a given stationary process, it is possible to identify the constituent frequencies through a sample-based estimate of the PSD/CPSD functions, also known as a periodogram. The spectral analysis decomposes the signal process into a sum of sine waves, which allows the assessment of its frequency content. Nonetheless, a sample-based estimate usually does not represent the underlying PSD/CPSD functions of a stochastic process, which is generally a challenge to define.

One of the methods for obtaining the underlying PSD/CPSD functions of a given signal process is through the ensemble average of a large set of repeated measurements. However, when it comes to wind tunnel testing, typical measurements are usually conducted just once or occasionally repeated a few times for a short time period, i.e., typically 30-60 seconds. In general, these individual records do not contain enough information to fully characterize the stochastic process. However, through smoothing techniques, it is possible to derive approximate PSD/CPSD functions of the process from a single record \citep{solomon1991psd}.  

The averaged Welch's method can be adopted to obtain the smoothed data-driven input PSD/CPSD functions \citep{Welch}. For a given set of zero-mean stationary stochastic processes $\mathbf{P}(t)$, the method consists of breaking the filtered signals into $K$ windowed segments of the same length and averaging their periodograms \citep{Welch}. This method reduces the variance of the PSD/CPSD functions through smoothing. The main objective of smoothing the PSD/CPSD functions is to reduce the noise (random effects associated with record-to-record variability), hence facilitating the identification of important patterns and trends that better align with the underlying PSD/CPSD functions of the process rather than the individual record. 

Let the $k$th windowed segment from the record, $P_i$, be denoted by:
\begin{equation} \label{signal_window}
{{P^{'}}_i}(m) = w(m)P_i(m+k Q)
\end{equation}
where $k=1,...,K$ with $K$ the number of segments, $Q$ is the number of data points to shift between the segments, $m=1,...,M$ with $M$ the number of data points in each segment, ${P^{'}}_i(m)$ is the windowed segment, and $w(m)$ is the window function (e.g., rectangular or Hanning). It should be noted that zero-padding the signal prior to dividing it up into segments may be necessary to ensure that $M$ corresponds to the frequency resolution defined by the predetermined time step and duration of the realization of Eq. (\ref{Eq_subprocess}). The cross-correlation matrix, $R_{{P}^{'}_{ij}}$, and CPSD matrix, $S_{P^{'}_{ij}}$, of a given pair of windowed signal segments, ${P}^{'}_i(m)$ and ${P}^{'}_j(m)$ with $i=1...N$ and $j=1...N$, are related through the Wiener-Khintchine transformation:
\begin{equation} \label{dft_seg}
S_{{P}^{'}_{ij}}(\omega) = \frac{1}{2\pi}\sum_{\tau=- \infty}^{+ \infty} R_{{P}^{'}_{ij}}(\tau)e^{- \textnormal{j}\omega \tau}
\end{equation}
where $\tau$ is the time lag. The principal diagonal of the CPSD matrix, $S_{P^{'}_{ij}}$, represents the auto-spectra, or PSD, of each component while the off-diagonals are the CPSD between a pair of components. The PSD functions are real and nonnegative, while the CPSD functions are generally complex.

The modified periodogram, $D_{k}(\omega)$, is obtained as:
\begin{equation} \label{mod_period}
D_{k}(\omega) = \frac{1}{W} |S_{P^{'}_{ij}}|^2
\end{equation}
where
\begin{equation} 
W=\sum_{m=0}^M w^2(m)
\end{equation}

By averaging the modified periodogram values by the number of total segments $K$, Welch's averaged PSD/CPSD functions are obtained as:
\begin{equation} \label{Savg}
\tilde{S}_{P^{'}_{ij}}(\omega) =\frac{1}{K} \sum_{k=1}^{K} D_{k}(\omega)
\end{equation}

The smoothed PSD/CPSD functions can then be used as input to calibrate the eigenvalues and eigenvectors of the POD-based stochastic wind model of Section \ref{Sect_DataDrivenPOD}.

%-------------------------------------------

\subsubsection{Target PSD/CPSD Functions} 
\label{wttarget}

While the smoothed PSD/CPSD functions obtained from a typical wind tunnel record can be used to calibrate the stochastic wind model, in an ideal case where many wind tunnel records are available, the target, or underlying, PSD/CPSD Functions can be obtained through the ensemble average as:
\begin{equation} \label{Savg_target}
\tilde{S}_{P_{ij,T}}(\omega) =\lim_{l\to \infty} \frac{1}{L} \sum_{l=1}^{L} S_{P_{ij,l}}(\omega)
\end{equation}
where $L$ is the total number of wind tunnel records. The averaged Welch's method can also be used to estimate the target PSD/CPSD functions of the stochastic process as long as a rectangular window and no overlapping are considered such that the segments and their frequency content are not affected by the end smoothing.

The knowledge of the target PSD/CPSD functions is necessary for this study to enable the quantification of errors associated with the use of the data-driven POD-stochastic wind load model calibrated to a typical wind tunnel record. 

%--------------------------------------------------------------
%--------------------------------------------------------------

\section{Error Quantification Approach}
\label{error_measures}

 \subsection{Overview}
This section describes the error measures adopted in this study. In the following, the term ``typical" spectra refer to the PSD/CPSD functions obtained from typical short-duration wind tunnel records. The ``target" spectra refer to the PSD/CPSD functions obtained through the ensemble average of multiple wind tunnel records, as described in Section \ref{wttarget}, and is used as the baseline for comparison. The term ``simulated" spectra refer to the PSD/CPSD functions obtained from realizations of the wind loads using the POD-based stochastic wind load model considering all modes, whereas the term ``approximated" spectra refer to the PSD/CPSD functions obtained from considering mode truncation. 

The wind tunnel variability is quantified by comparing typical spectra to the target spectra. The errors associated with the stochastic wind load model are quantified by estimating both the model errors (using large enough sample sizes to render statistical errors negligible) and mode truncation errors, i.e., the difference between the target and the approximated spectra obtained by simulation with higher modes truncated.

 %----------------------------------------------------------------
 
 \subsection{Proposed Error Measures}

The errors associated with typical wind tunnel records, model errors, and errors introduced by mode truncation are quantified using the error measures described in this section, where the target spectra are considered as the reference quantity value for all estimated error measures. In general, the errors between a given target spectra and typical spectra are quantified in terms of the integration of the PSD/CPSD functions in the frequency domain. The integral of the PSD function will provide an estimate of the variance of the signal while the integral of the cospectrum $(Co_{ij})$, the real part of the CPSD function, will provide an estimate of the covariance between the two signals \citep{huppenkothen2018statistical}. The variance, $\sigma^2_{ii}$, and covariance, $\sigma_{ij}$, are therefore estimated as:
\begin{equation} \label{variance}
\sigma^2_{ii} = \int_{-\infty}^{+\infty} {S}_{P_i}(\omega) d\omega
\end{equation}
\begin{equation} \label{covariance}
\sigma_{ij} = \int_{-\infty}^{+\infty} Co_{ij}(\omega) d\omega
\end{equation}
where $i=1,...N$ and $j=1,...,N$ denote the $i$th and $j$th components of the vector-valued process $\textbf{P}(t)$, respectively. If filtering is used to smooth out or reduce signal noise, the variance and covariance are then obtained by integrating the PSD/CPSD functions up to the filtering frequency. 

To quantify the error due to wind tunnel record variability, the error in variance between the target spectra and typical spectra is defined as:
\begin{equation} \label{error_area_auto}
 \varepsilon_i(\%) =\left( \frac{\sigma_{ii}^2-\sigma_{ii_T}^2}{\sigma_{ii_T}^2}\right)\times 100 
\end{equation}
where $\sigma_{ii}^2$ is the variance obtained from the typical PSD functions while $\sigma_{ii_T}^2$ is the variance obtained from the target PSD functions.

Percentage error estimates for pairs in the off-diagonal and poorly correlated signals are difficult to obtain with precision and may have an unacceptably large error. The target covariance, $\sigma_{ij_T}$, can have values close to zero for signals with weak correlation, therefore, leading to large relative errors when $\sigma_{ij_T}$ is used as the denominator in the percentage error estimation. Therefore, in this study, errors in the CPSD functions are assessed by calculating the difference between the correlation coefficients of the target spectra and the correlation coefficients of each typical spectra, and therefore as:
\begin{equation} \label{corrcoef_diftarget}
 \rho_{t,ij}= \frac{\sigma_{ij_T}}{\sqrt{\sigma^2_{ii_T}\sigma^2_{jj_T}}}
\end{equation}
\begin{equation} \label{corrcoef_difreal}
 \rho_{r,ij}= \frac{\sigma_{ij_r}}{\sqrt{\sigma_{ii_r}^2\sigma_{jj_r}^2}}
\end{equation}
\begin{equation} \label{corrcoef_dif}
 \varphi_{ij}=\rho_{t,ij}-\rho_{r,ij}
\end{equation}
where $\rho_{t,ij}$ is the correlation coefficient of the target spectra, $\rho_{r,ij}$ is the correlation coefficient of each typical spectra, $r=1,..., R$ with $R$ the total number of wind tunnel records, and $\varphi_{ij}$ is the difference between the correlation coefficients of the target and typical spectra. As this assessment is in terms of correlation coefficients, which range from -1 to 1, it is straightforward to judge/interpret the significance of any observed differences. It is important to mention that the measures of variance and correlation coefficients are assumed sufficient to describe the stochastic process and estimate the associated errors of the Gaussian wind load process.

To estimate the model and mode truncation errors, the same error measures described in Eq. (\ref{error_area_auto})-(\ref{corrcoef_dif}) are adopted where the ``typical" spectra are replaced by the ``simulated" and ``approximated" spectra, respectively. It should be noted that to obtain errors purely from the stochastic simulation model without the influence of record variability, the target spectra are used as input to calibrate the stochastic wind model to generate the ``simulated" and ``approximated" realizations of the wind load process for error quantification.

%--------------------------------------------------------------
%--------------------------------------------------------------

\section{Experimental Setting} \label{experimental}

\begin{figure}
	\centering
		\includegraphics[scale=.5]{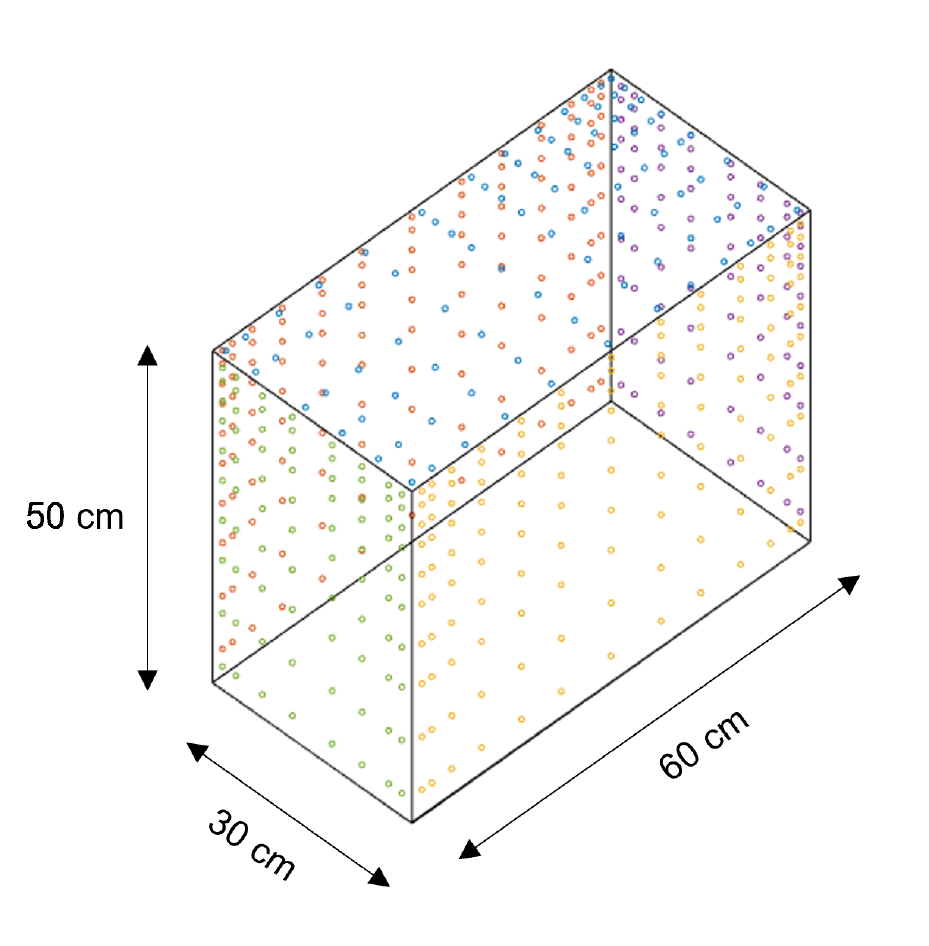}
	\caption{Illustration of the building model used in the wind tunnel tests and configuration of the pressure taps.}
	\label{model_taps}
\end{figure}

\begin{figure*}
    \centering
    \includegraphics[width=0.75\textwidth]{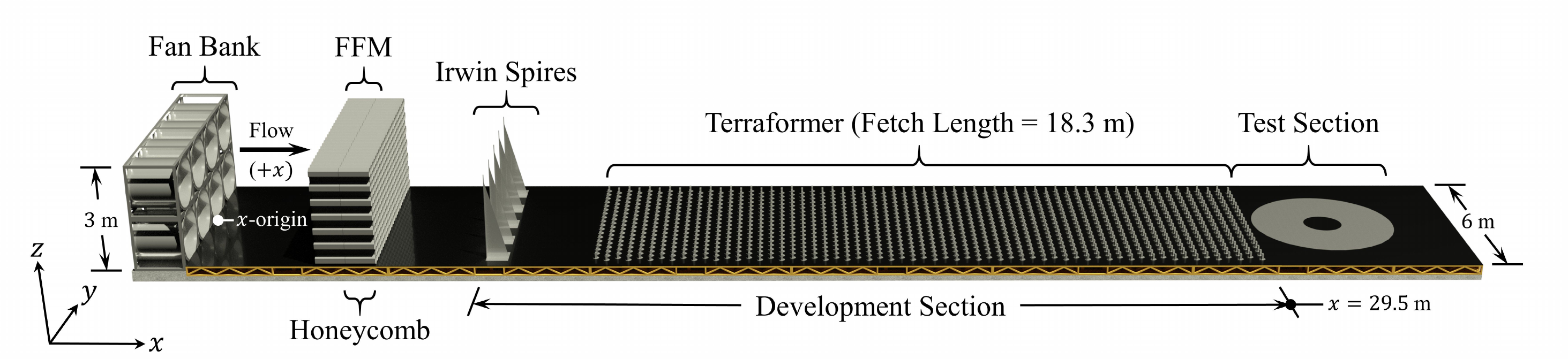}
    \caption{Representation of the UF NHERI boundary layer wind tunnel.}
    \label{tunnelprofile}
\end{figure*}

To enable the error quantification of Section \ref{error_measures}, extensive experimental testing was carried out. A rectangular rigid model was employed in this study, with geometry and configuration of the pressure taps shown in Fig. \ref{model_taps}. Data on the pressure acting on the model envelope was simultaneously collected using a Scanivalve system and pressure taps on the surface of the building model. The model had 512 pressure taps in total, distributed on five surfaces to make sure that the flow around the corners and flow variation with height, width, and depth were captured. The experiments were carried out at the Natural Hazards Engineering Research Infrastructure (NHERI) Boundary Layer Wind Tunnel of the University of Florida. The tunnel is 6 m wide, 3 m tall, and 40 m long, as shown in Fig. \ref{tunnelprofile} where the “terraformers” are automated terrain roughness elements with an adjustable height that can be quickly adjusted by electric actuators \citep{catarelli2020automation}. For the conducted tests, the height of the terraformer elements was set to 16 cm, in order to obtain a suburban terrain condition. The model scale considered in this study was 1:200, representing a 25-story full-scale building.

The external pressure coefficients, $C_{p,e}$, at each tap location and time, $t$, are estimated as:
\begin{equation} \label{presscoef}
C_{p,e}(t)=\frac{p(t)-p_0}{q}
\end{equation}
where $p(t)$ is the raw pressure measured at each pressure tap, $p_0$ is the mean reference static pressure measured by a Cobra Probe located upwind and at the height of the model, and $q$ is the dynamic pressure defined as $q=1/2\rho U_H^2$ with $\rho$ the local air density and $U_H$ the mean streamwise wind speed at the model height. Information on the local atmospheric pressure and temperature was used to estimate the air density during the experiments.

\begin{figure}
	\centering
		\includegraphics[scale=.6]{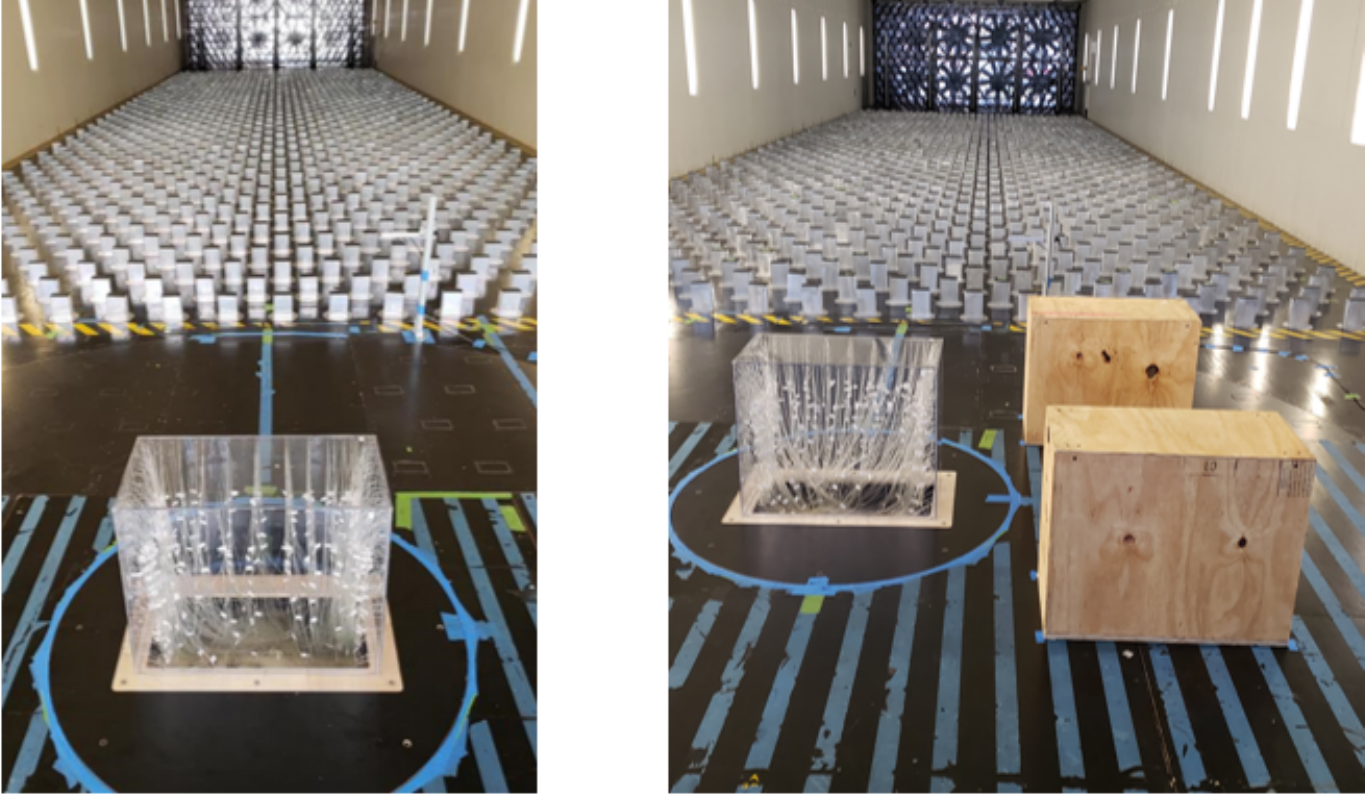}
	\caption{Rectangular building model used in the wind tunnel tests, with SM setup shown on the left, and PM setup shown on the right, for a wind direction of $\beta = 0$ degrees.}
	\label{model_tunnel}
\end{figure}

\begin{figure}
	\centering
		\includegraphics[scale=.6]{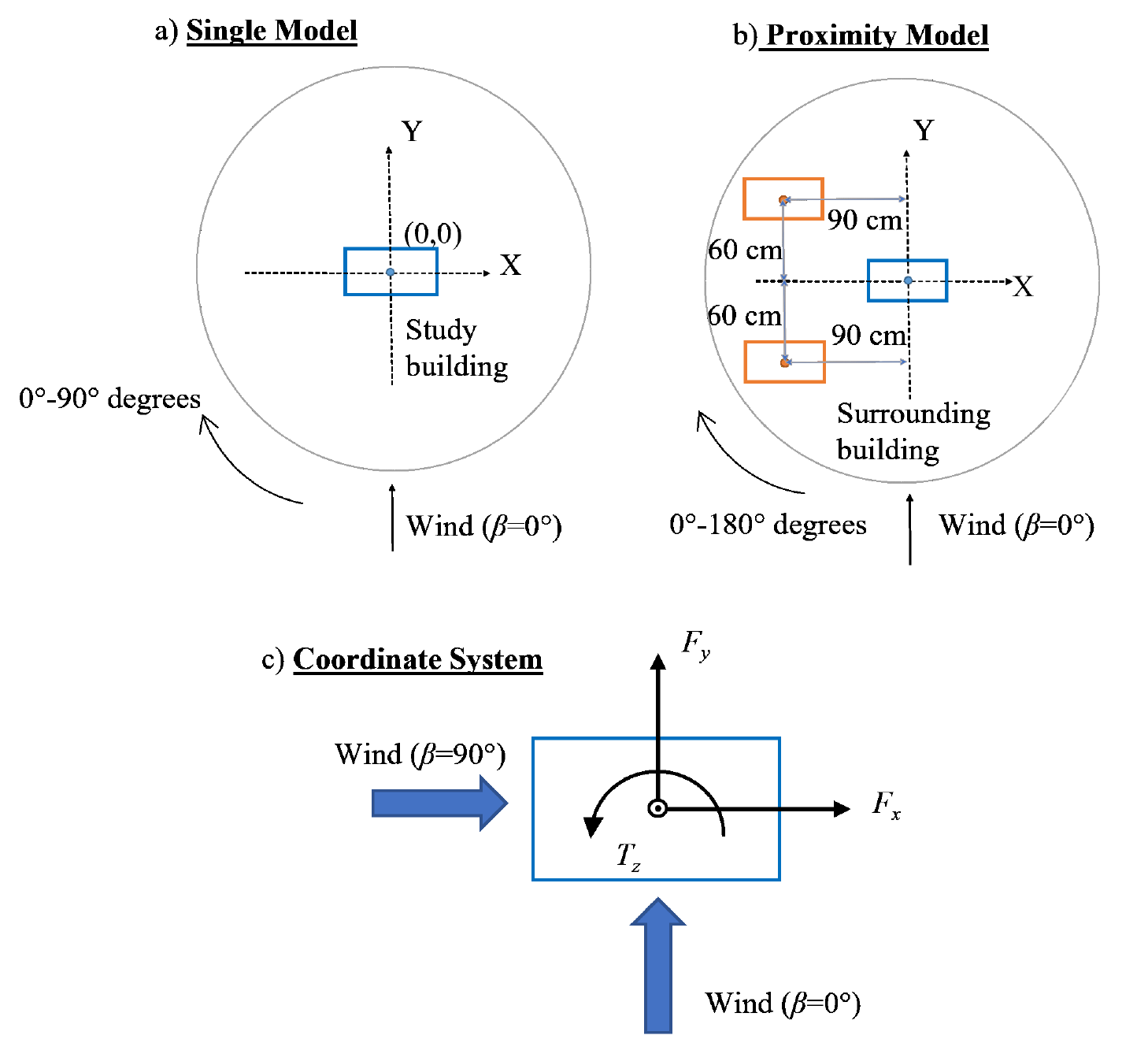}
	\caption{(a) SM setup, (b) PM setup, and (c) coordinate system adopted in estimating the wind loads.}
	\label{testing_scheme}
\end{figure}

In order to investigate possible interference effects on the errors associated with the stochastic wind load model, two different testing configurations were considered. The first considered a single model (SM) located at the center of the wind tunnel testing section of Fig. \ref{tunnelprofile}, while the other configuration added two proximity models (PM) with the same geometry as the building model, as shown in Fig. \ref{model_tunnel}.  For the SM setup, the experiment was conducted for approximately 15 minutes and repeated five times for, due to the symmetry, wind directions varying from 0 to 90 degrees, in increments of 10 degrees, as shown in Fig. \ref{testing_scheme}(a). Due to the bilateral symmetry of the proximity model test setup, as shown in Fig. \ref{testing_scheme}(b), five repetitions of 15 minutes were carried out rotating from 0 to 180 degrees, also in 10 degrees increments. The data was collected with a sampling frequency of 625 Hz. 

The forces acting on the $n$th floor in the $x$ and $y$ translational directions as well as rotation around the $z$ direction were obtained through the integration of linearly-interpolated pressures over the tributary area corresponding to each floor of the building. The resultant forces and moment can be decomposed into $F_{x,n}(t)$, $F_{y,n}(t)$ and $T_{z,n}(t)$ components, where $n$ denotes the floor number. Figure \ref{testing_scheme}(c) shows the forces, coordinate system, and wind directions.
The force components at each floor are normalized to obtain force coefficients as follows:
\begin{equation} \label{force_coeff}
CF_{x,n}(t) = \frac{F_{x,n}(t)}{q B_x H}
\end{equation}
\begin{equation}
CF_{y,n}(t)  = \frac{F_{y,n}(t)}{q B_y H}
\end{equation}
\begin{equation}
CT_{z,n}(t) =      \frac{T_{z,n}(t)}{q  {H}\frac{B_{\max}^2}{2}}
\end{equation}
where $CF_{x,n}(t)$, $CF_{y,n}(t)$, $CT_{z,n}(t)$  are the dynamic force coefficients at the $n$th floor, $H$ is the height of the model, $B_x$ and $B_y$ are the plan dimensions of the building, and $B_{\max}$ is the maximum plan dimension of the building.

%-----------------------------------------

\subsection{Assessment of Wind Tunnel Data} 
\label{targetcpsd}
In order to evaluate the error associated with using typical wind tunnel records, as defined in Section \ref{error_measures}, the target spectra and the typical spectra are needed. To obtain those spectra, the wind tunnel data set was divided into two groups; the first group was used to define the target spectra and the second group was used to obtain the typical spectra, also referred to as the testing set. The purpose of dividing the data into two different groups is to obtain unbiased and meaningful error estimates as the testing set is independent of the data set used to define the target.

The first 10 minutes of each repetition were used to define the target spectra through Welch’s averaged method, as described in Section \ref{wttarget}. In particular, each 10-minute set was divided into 150 segments of 4-second duration, while a rectangular window of the same length as the signal was used to ensure zero padding and zero overlapping between the segments. The window size of 4 seconds was adopted since it was small enough to obtain multiple periodograms to be averaged out to obtain a smooth curve as well as to capture the low-frequency content sufficiently well without any padding-based interpolation. A total of 750 segments were used to establish the target spectra for each force coefficient. Ultimately, the target spectra were estimated for all 75 force components for every wind direction and setting. 

The remaining data were used as a testing set, which was compared to the target to evaluate the errors as described in Section \ref{error_measures}. The testing data were divided into typical 32-second segments, which were treated as independent wind tunnel records. The typical spectra of each segment were obtained by splitting the segment into $K$ overlapping blocks and applying a window function as described in Section \ref{spectra_welch}. The Hanning window function with an overlap of 50\% was adopted. In general, a total of 45 records were obtained for each wind direction, with a few directions having fewer records (e.g., 35-44 segments) due to missing data or evident errors in the data set (e.g., abnormal peak pressure disturbance), hence they were disregarded.

As only frequencies up to a certain value are of interest in defining spectra for practical wind engineering applications, a second-order Butterworth lowpass filter with a cutoff frequency of 50 Hz was considered when defining the target and typical spectra. The cutoff frequency was determined such that unrealistic high-frequency noise caused by equipment was eliminated. Therefore, the integration limits of the spectra, as shown in Eqs. (\ref{variance}) and (\ref{covariance}), was defined by the cutoff frequency of the filter.

%--------------------------------------------------------------
%--------------------------------------------------------------

\section{Results}
\subsection{Overview}

The wind tunnel data collected as described in Section \ref{experimental} is first employed to define the target spectra to enable the assessment of the errors induced by the variability in the typical spectra estimated from single wind tunnel records. The target spectra are then used to calibrate the data-driven stochastic wind load model and simulate the wind process realizations, which are subsequently used to quantify the model errors. %The results for the errors associated with typical wind tunnel records, statistical errors, and errors introduced by mode truncation are presented in this section.

\subsection{Wind Tunnel Data Variability}

\begin{figure}
	\centering
		\includegraphics[scale=.8]{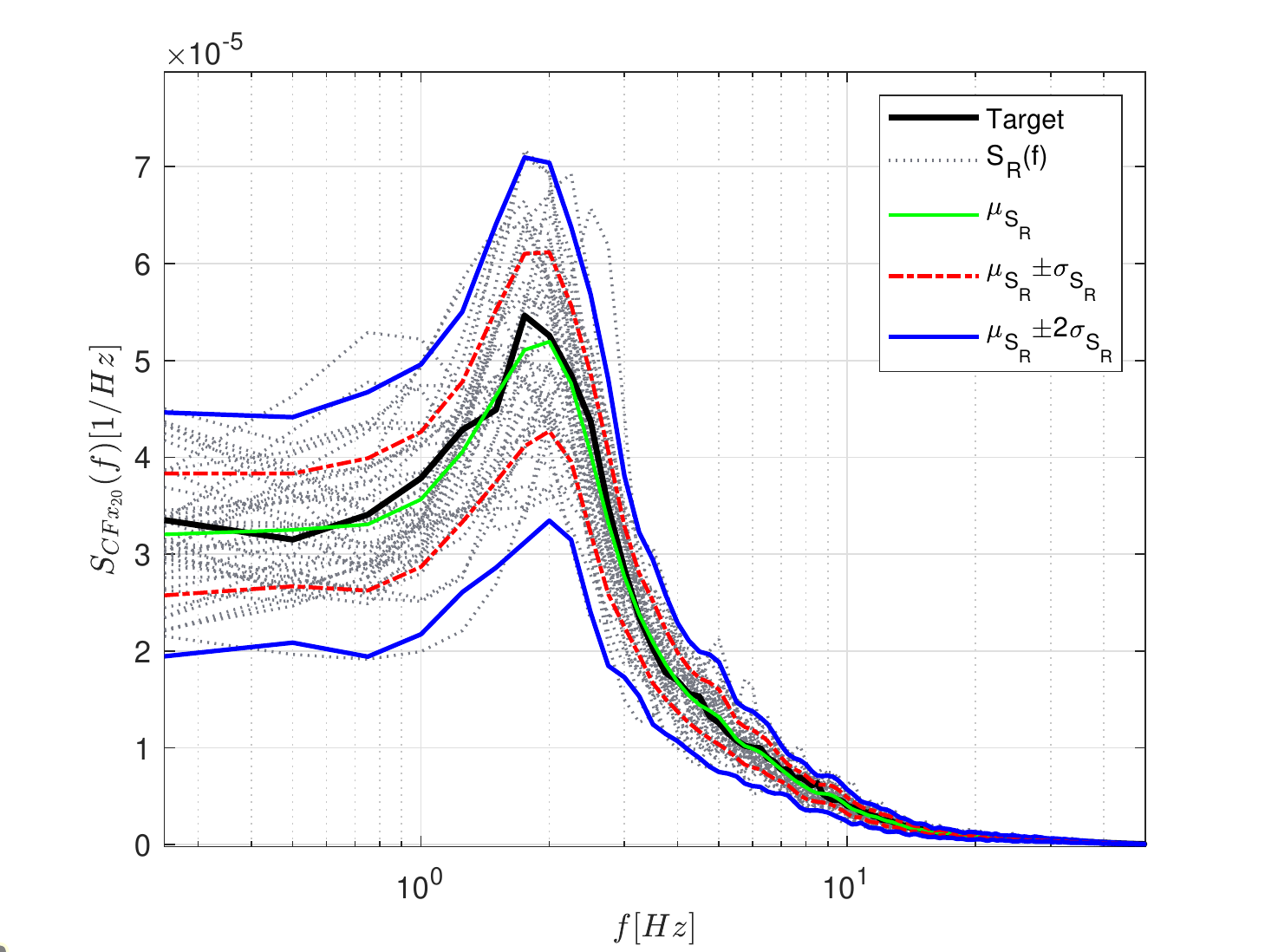}
	\caption{Comparison between the typical and target PSD of $CF_{x,20}(t)$ at the 20th floor and a wind direction of $\beta=0^{\circ}$.}
	\label{spectra11}
\end{figure}

\begin{figure}
	\centering
		\includegraphics[scale=.8]{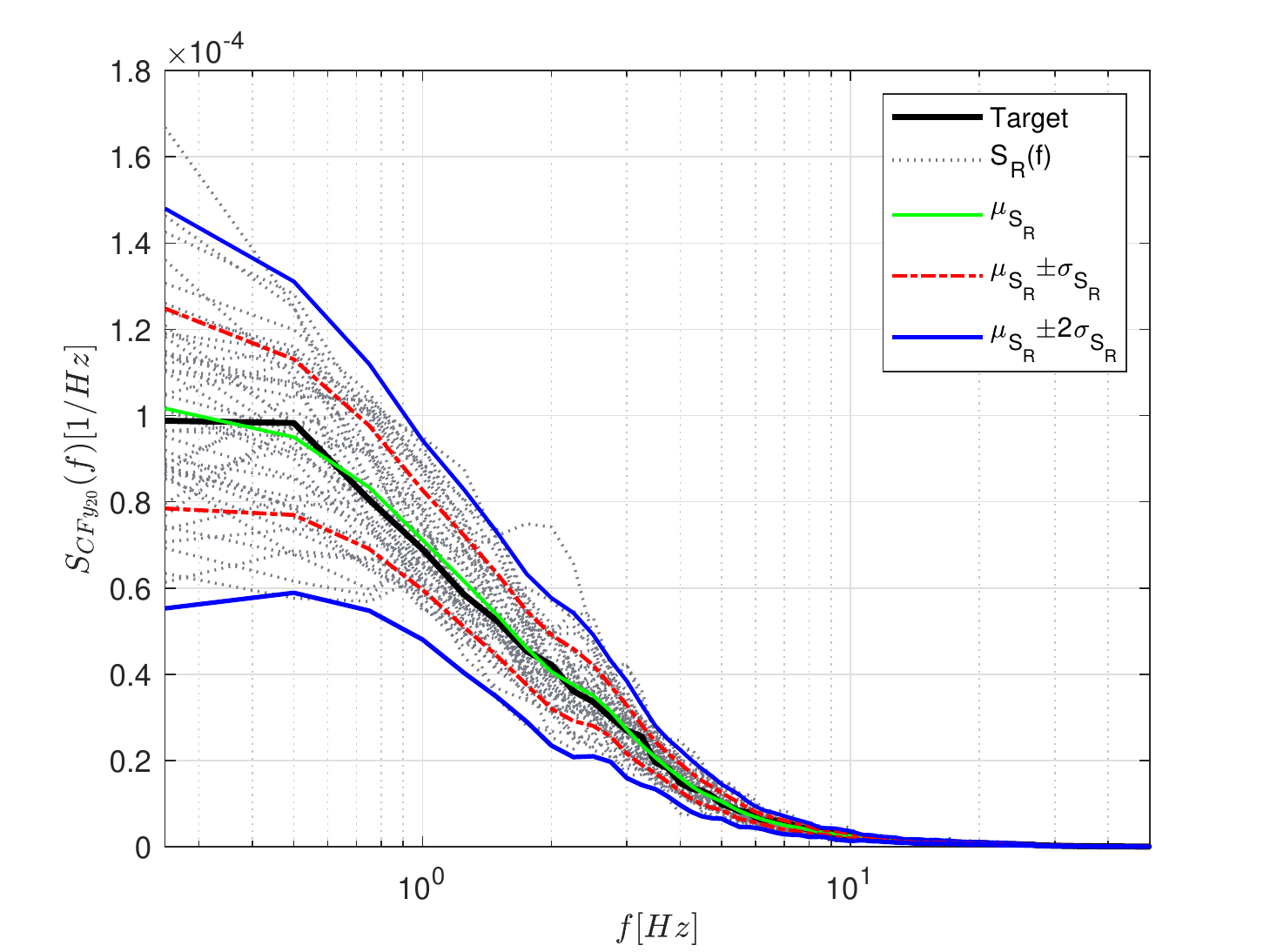}
	\caption{Comparison between the typical and target PSD of $CF_{y,20}(t)$ at the 20th floor and a wind direction of $\beta=0^{\circ}$.}
	\label{spectra22}
\end{figure}

\begin{figure}
	\centering
		\includegraphics[scale=.8]{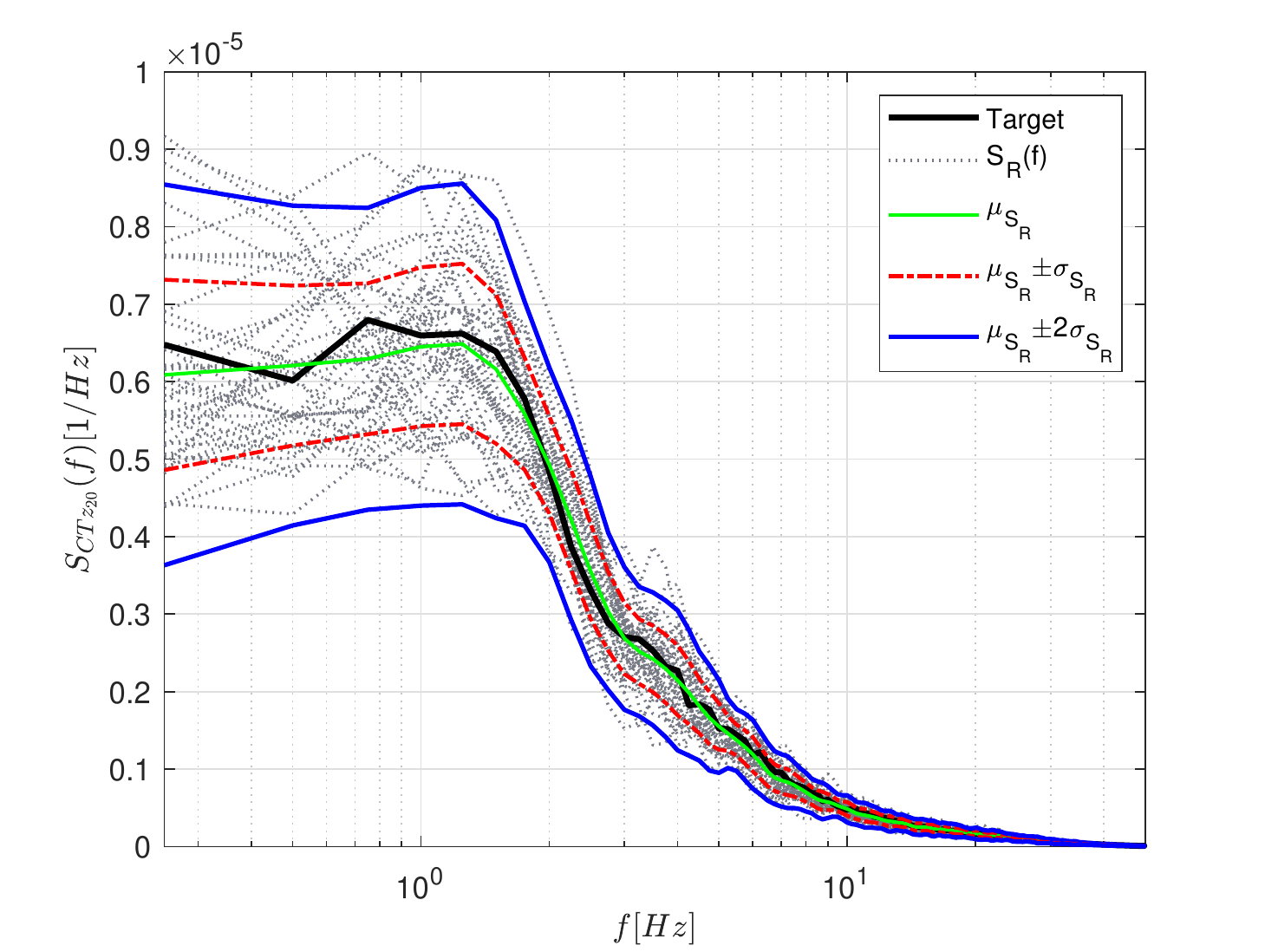}
	\caption{Comparison between the typical and target PSD of $CT_{z,20}(t)$ at the 20th floor and a wind direction of $\beta=0^{\circ}$.}
	\label{cospectra21}
\end{figure}

The variability induced by the use of typical wind tunnel records is assessed by comparing the typical spectra (estimated for a 32-second wind tunnel record) to the target spectra in terms of the variance and correlation coefficients. Figures \ref{spectra11}-\ref{cospectra21} provide a comparison between the target PSDs and the typical PSDs, $S_R$, associated with force coefficients at the $20$th floor, i.e. $CF_{x,20}(t)$, $CF_{y,20}(t)$ and $CT_{z,20}(t)$, respectively. As can be observed, the typical PSDs clearly fluctuate around the target spectra, showing record-to-record variability up to two standard deviations from the mean. However, $\mu_{S_R}$, which is the mean of the typical spectra, is close to the target spectra for all $CF_{x,20}(t)$, $CF_{y,20}(t)$ and $CT_{z,20}(t)$ components. This indicates the quality and repeatability of the experiment. A similar conclusion could be drawn for other floors.

\begin{figure}
	\centering
		\includegraphics[scale=.8]{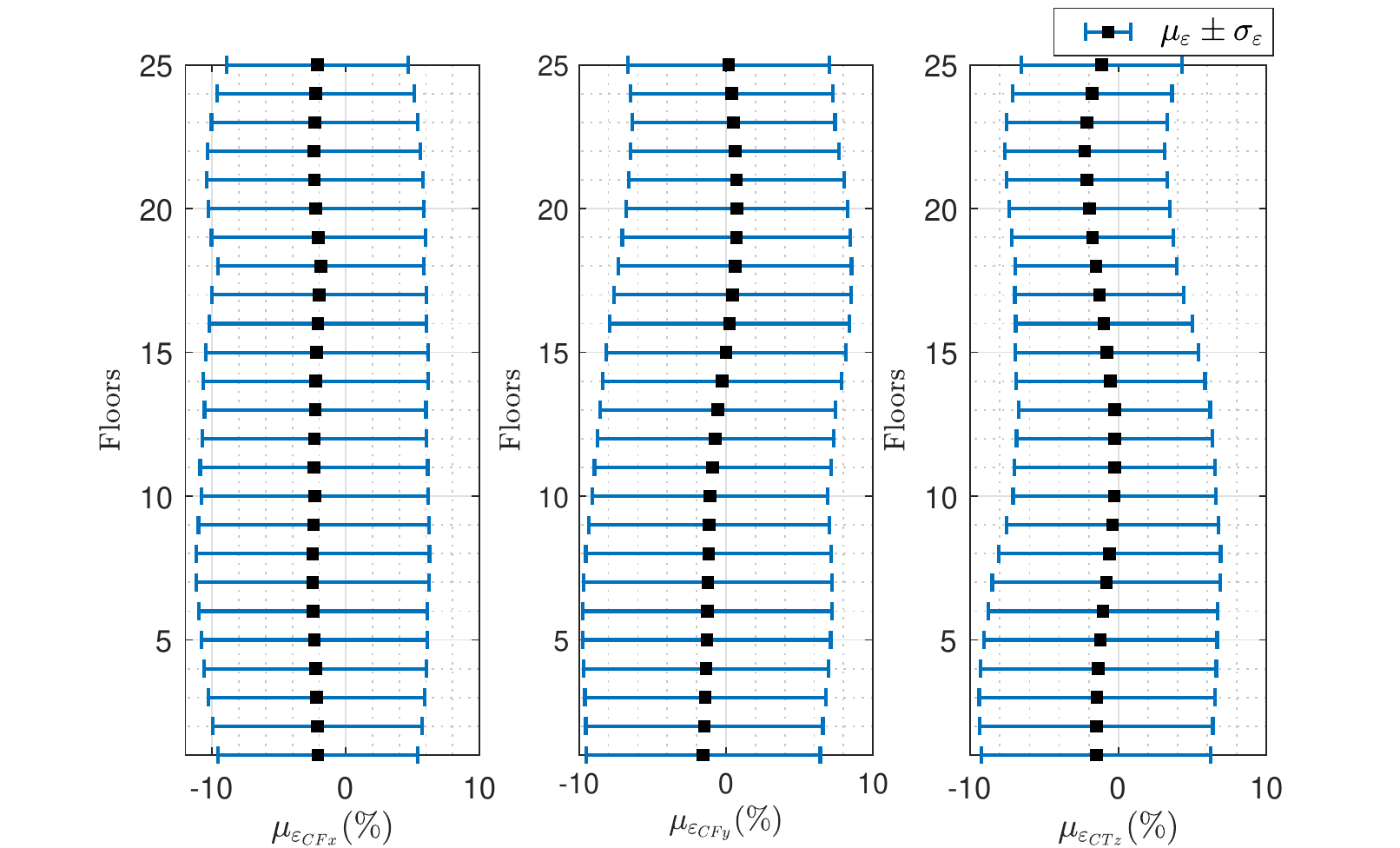}
	\caption{Mean error in the variance, $\mu_{\varepsilon}$, with standard deviation, $\sigma_{\varepsilon}$, for each floor, for the SM layout and $\beta=0^{\circ}$.}
	\label{ang000-CFxyT}
\end{figure}

\begin{figure}
	\centering
		\includegraphics[scale=.8, angle=90]{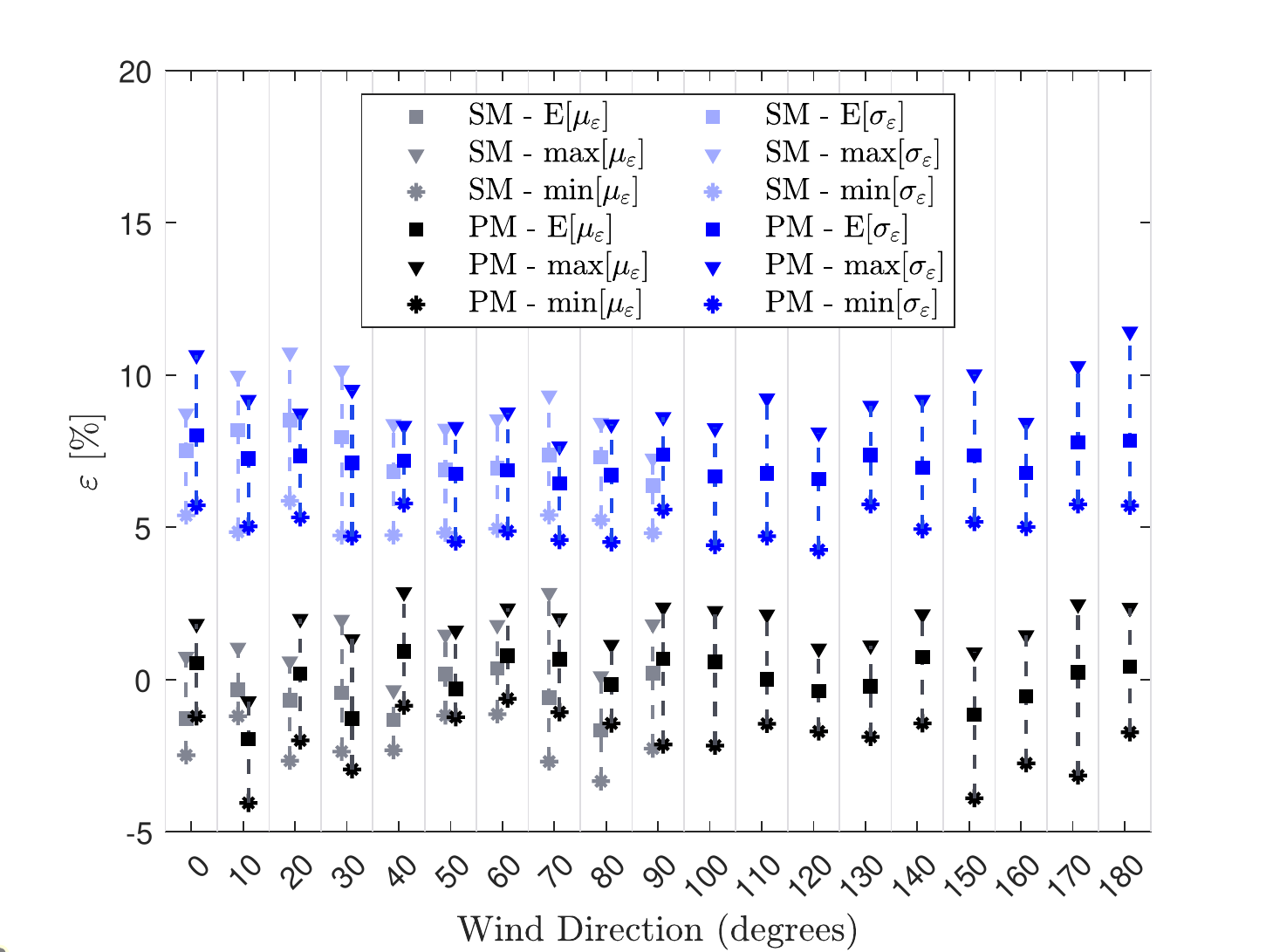}
	\caption{Expected values and range of the mean and standard deviation of the error in variance for both the SM and PM setups.}
	\label{meanerror_alldir}
\end{figure}

Figure \ref{ang000-CFxyT} presents the mean, $\mu_{\varepsilon}$, and standard deviation, $\sigma_{\varepsilon}$, of the error in the variance over all records for all force components at each floor for the SM layout and wind direction $\beta=0^{\circ}$. It is evident that $\mu_{\varepsilon}$ ranges between -2.5\% to 0.8\% for all force components and all floors, while $\sigma_{\varepsilon}$ varies between 5.4\% to 8.7\%. The expectation, maximum, and minimum values of $\mu_{\varepsilon}$ and $\sigma_{\varepsilon}$ for all wind directions and both experimental configurations are summarized in Fig. \ref{meanerror_alldir}. The expectation of $\mu_{\varepsilon}$ over all floors and force components, E$[\mu_{\varepsilon}]$, ranges between -1.7\% to 0.4\% for the SM layout and between -2\% to 0.9\% for the PM layout. The range of maximum and minimum values of $\mu_{\varepsilon}$ is between -4\% to 2.9\% for all wind directions. Regarding the standard deviation $\sigma_{\varepsilon}$, the maximum and minimum values also presented a consistent range for all wind directions and experimental settings, with the expected value, E$[\sigma_{\varepsilon}]$, ranging from 6.4\% to 8.5\%. This implies that the use of typical wind tunnel records can cause considerable deviation from the target independently of force component and wind direction. %This error can be explained by the inherent randomness of the data.  
Therefore, it can be inferred that this error is not negligible and can propagate through the simulation process. Consequently, this type of uncertainty needs to be accounted for when using stochastic wind load models calibrated to typical wind tunnel records. 

\begin{figure}
	\centering
		\includegraphics[scale=.7]{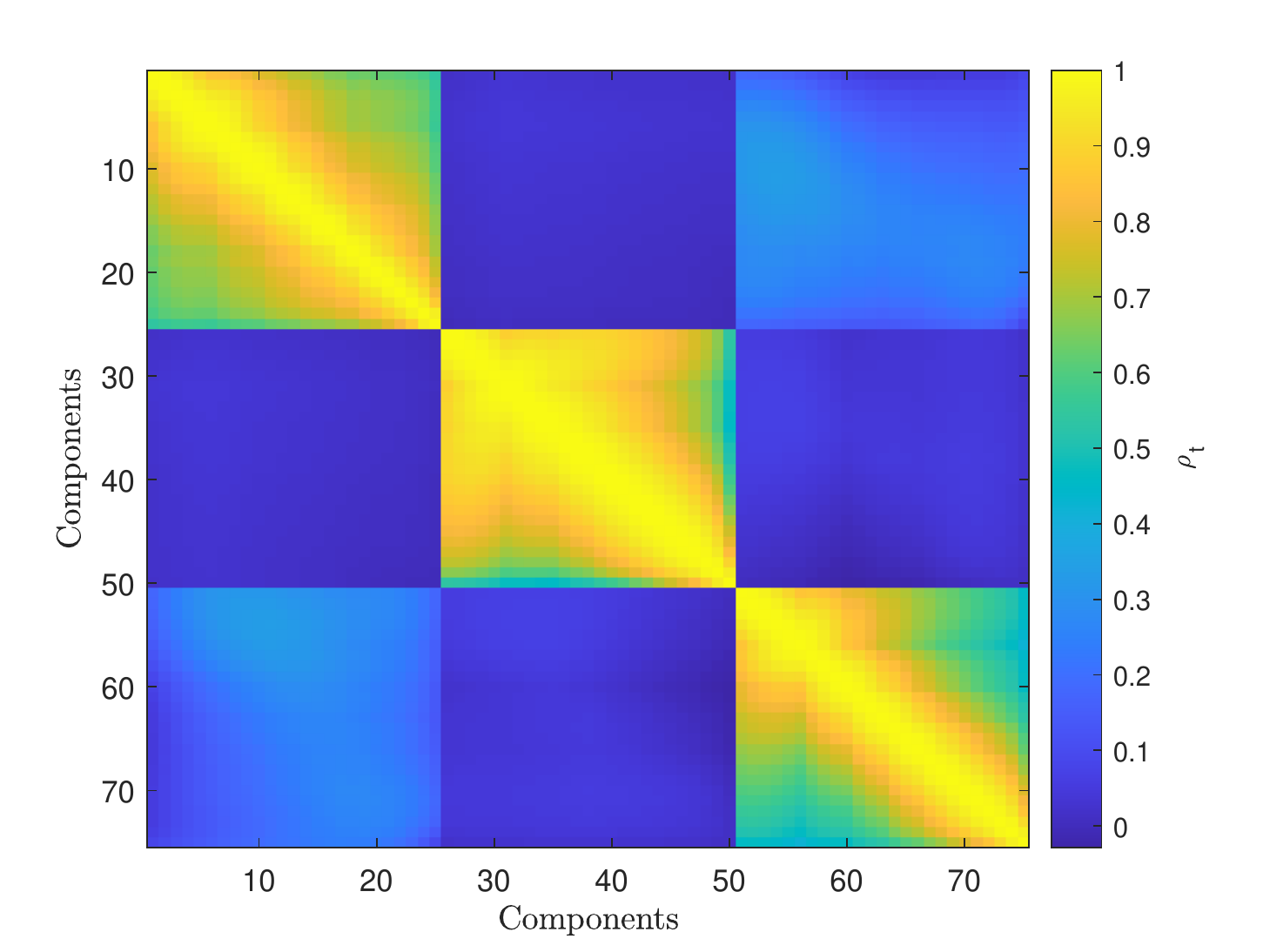}
	\caption{Map of the target correlation coefficients of the SM setup and wind direction $\beta=0^{\circ}$.}
	\label{corrcoef_target}
\end{figure}

\begin{figure}
	\centering
		\includegraphics[scale=.7]{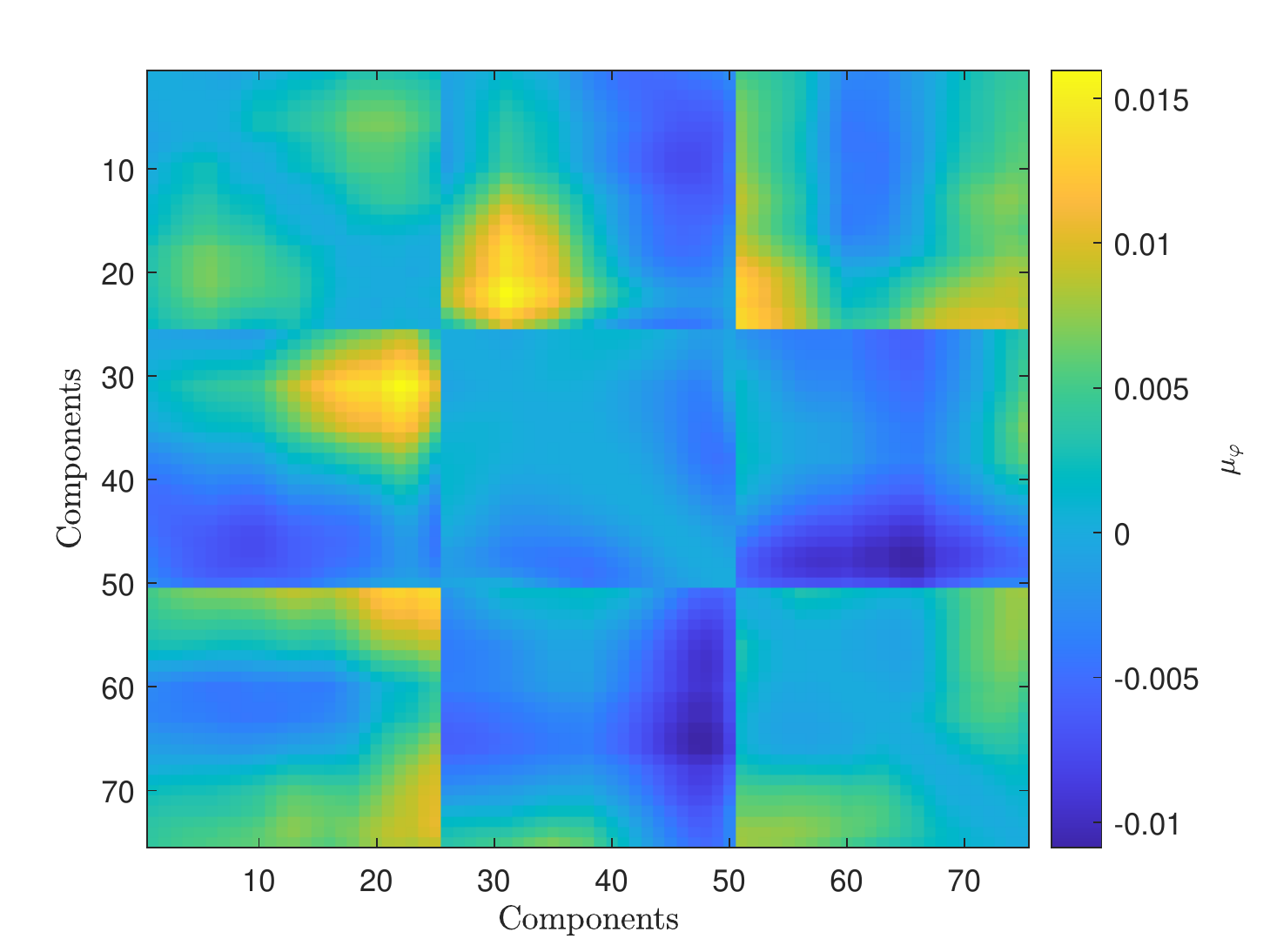}
	\caption{Mean value of the difference between target and typical correlation coefficients for the SM setup and wind direction $\beta=0^{\circ}$.}
	\label{diffcorrcoef_mean}
\end{figure}

\begin{figure}
	\centering
		\includegraphics[scale=.7]{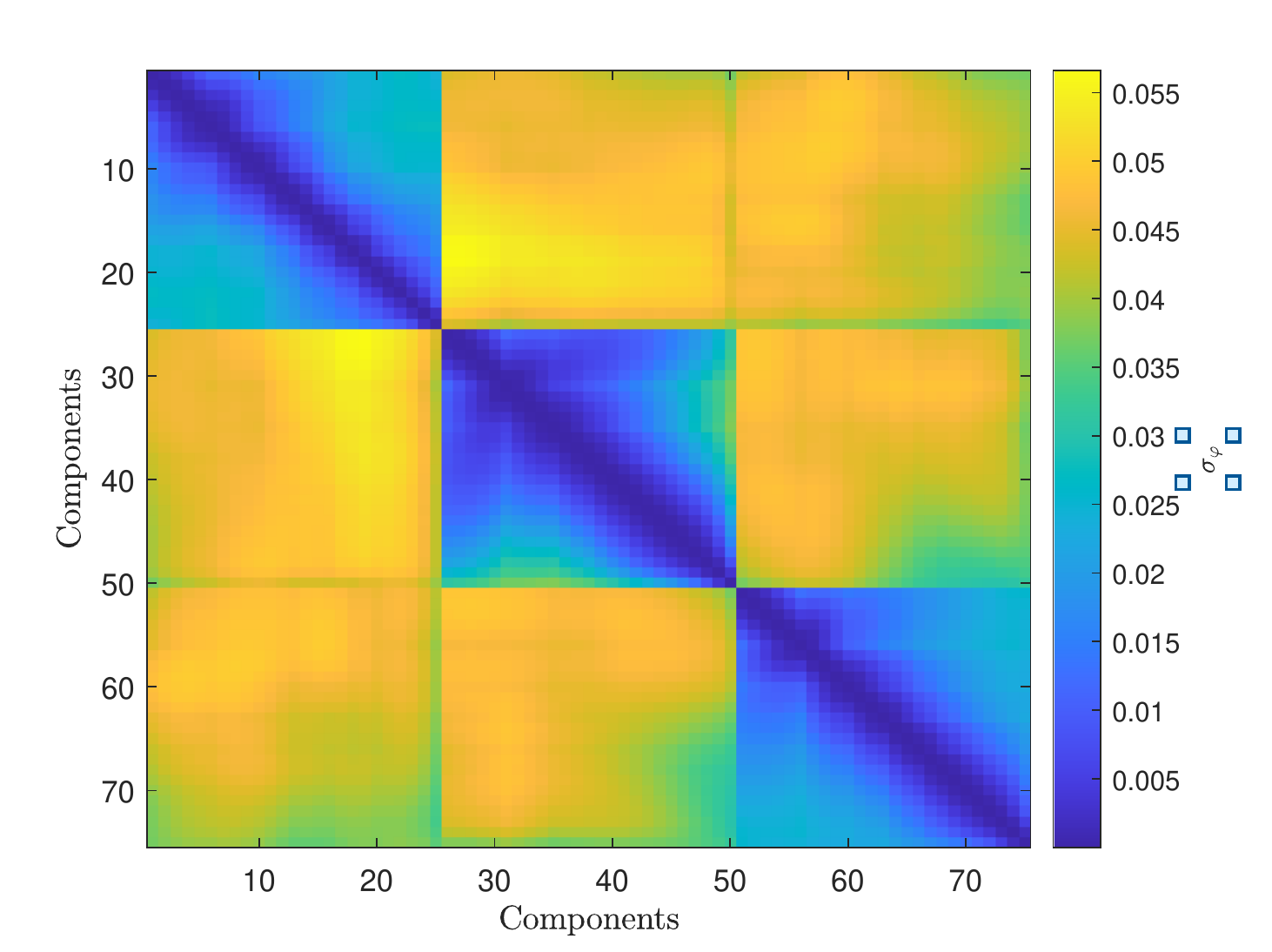}
	\caption{Standard deviation of the difference between target and typical correlation coefficients for the SM setup and wind direction $\beta=0^{\circ}$.}
	\label{diffcorrcoef_std}
\end{figure}

The correlation coefficients of the target spectra are presented in Fig. \ref{corrcoef_target} for $\beta=0^{\circ}$ and the SM setup. It should be noted that components 1-25, 26-50, and 51-75 refer to components of $CF_x(t)$, $CF_y(t)$, and $CT_z(t)$, respectively. It can be observed that a highly positive correlation occurs for components acting in the same direction. There is almost no correlation between $CF_x(t)$ and $CF_y(t)$, and $CF_y(t)$ and $CT_z(t)$, but there is a certain degree of positive correlation between $CF_x(t)$ and $CT_z(t)$. This result agrees with previously obtained experimental results for similar building geometries and wind direction of $\beta=0^{\circ}$ \citep{lin2005characteristics}. Figure \ref{diffcorrcoef_mean} illustrates the map of the mean, $\mu_{\varphi}$, of the difference between correlation coefficients, ${\varphi}$ obtained as defined in Eq. (\ref{corrcoef_dif}), obtained from the target and typical spectra for $\beta=0^{\circ}$. The mean difference ranges from -0.012 to 0.016, which is relatively small compared to the magnitude of the target correlation coefficients in Fig. \ref{corrcoef_target}, indicating the quality of repeatability of the experiment. Figure \ref{diffcorrcoef_std} shows the map of the standard deviation, $\sigma_{\varphi}$, of the difference in correlation coefficients for $\beta=0^{\circ}$ and the SM configuration. It can be observed that highly correlated components show a smaller variability, while the higher $\sigma_{\varphi}$ values are for those pairs with a very small correlation coefficient. The values of $\sigma_{\varphi}$ reach as high as 0.056, indicating uncertainty in the estimation of the correlation coefficients.

\begin{figure}
	\centering
		\includegraphics[scale=.8]{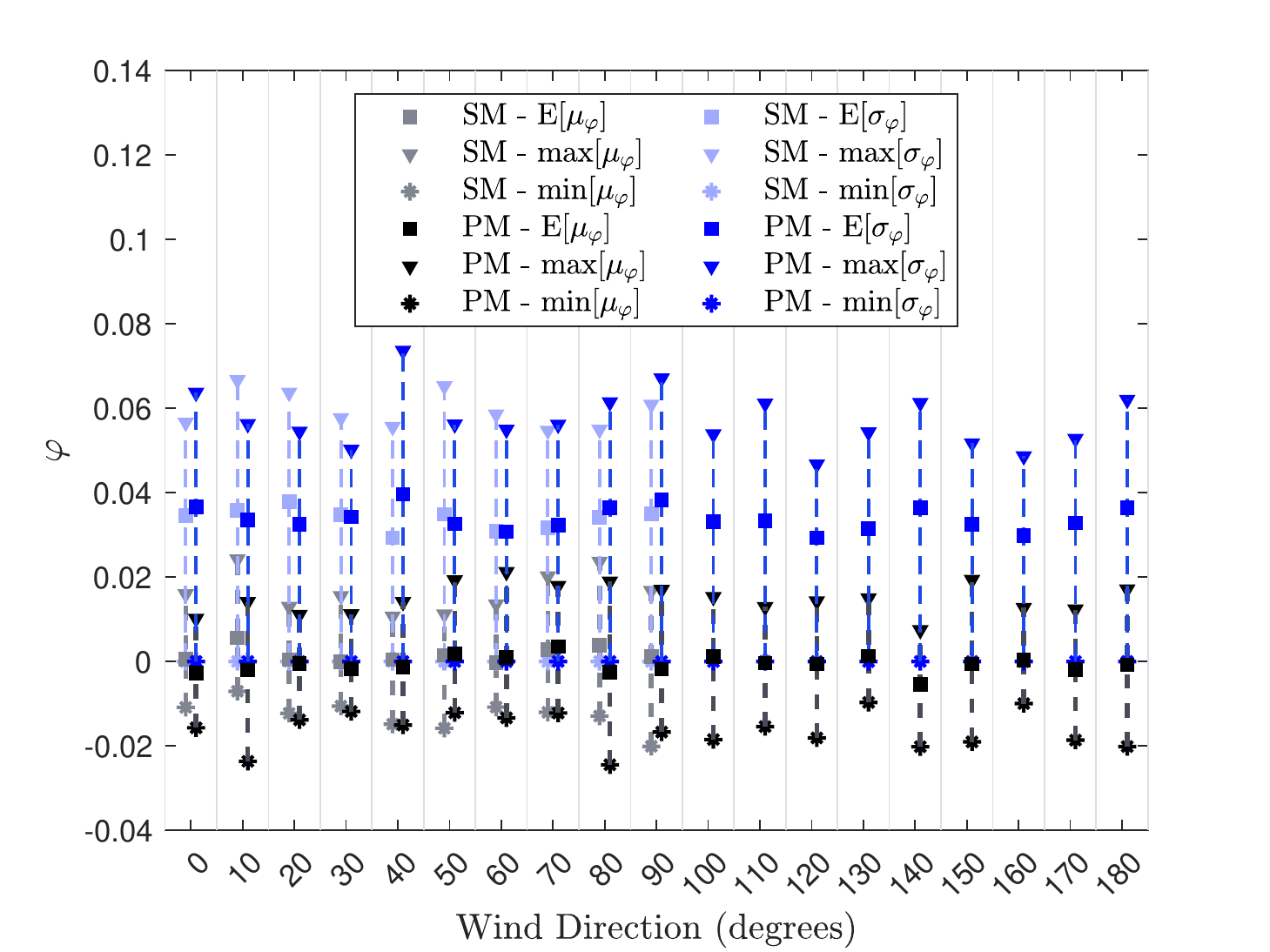}
	\caption{Expected values and range of the mean difference in the correlation coefficients for SM and PM layouts.}
	\label{meandiff_alldir}
\end{figure}

Figure \ref{meandiff_alldir} shows a summary of the previously discussed maps (Figs. \ref{diffcorrcoef_mean}-\ref{diffcorrcoef_std}) for all wind directions and both experimental configurations. The expectation of $\mu_{\varphi}$, E$[\mu_{\varphi}]$, over all floors and force components fluctuates around zero with a minimum of -0.022 and a maximum of 0.022. The small values of E$[\mu_{\varphi}]$ that fluctuate around zero indicate that on average the correlation structure estimated from the typical spectra is close to the target. The expected value of the standard deviation of $\varphi$, E$[\sigma_{\varphi}]$, fluctuates around 0.04 with a minimum of 0 and a maximum of 0.08, indicating some level of variability in the correlation structure. A similar trend is observed for all wind directions and settings.

\begin{figure}
    \centering
    \includegraphics[scale=0.9]{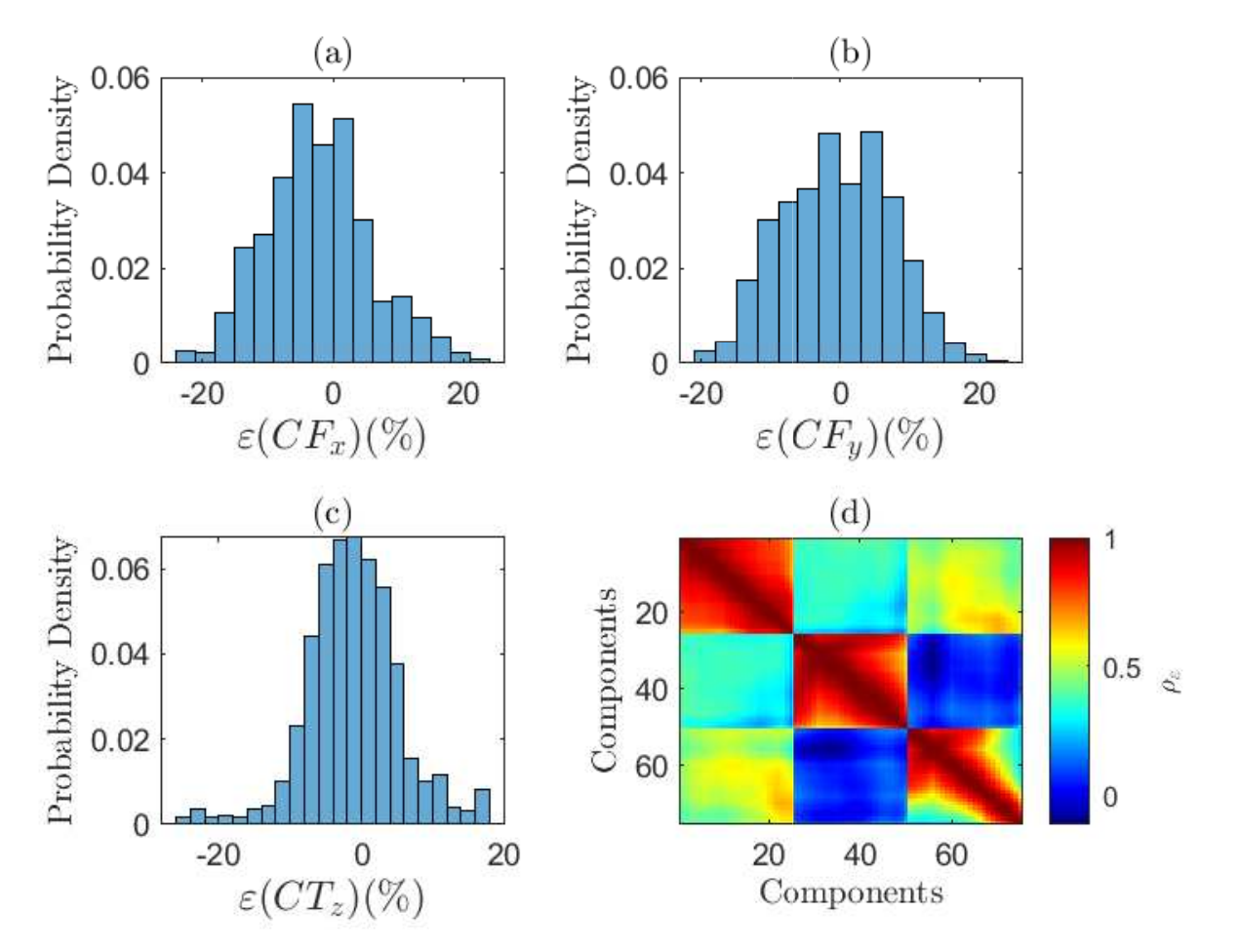}
    \caption{Histogram of $\varepsilon$ associated with (a) $CF_x$; (b) $CF_y$; and (c) $CT_z$, and (d) correlation coefficients between the components of $\varepsilon$.}
    \label{dist_corr_error}
\end{figure}

Figure \ref{dist_corr_error}(a)-(c) presents the histograms of the error in variance, $\varepsilon$, from all realizations in the testing set for the SM layout and $\beta=0^{\circ}$. The histograms resemble approximately normal distributions. The map of Fig. \ref{dist_corr_error}(d) shows the correlation of the error in variance, $\rho_{\varepsilon}$, associated with each force component. It can be observed that there is a high correlation in $\varepsilon$ between components acting in the same direction, but a poor or moderate correlation for components associated with different directions. A similar trend was observed for other directions and settings. This information can potentially be used to derive a factor that accounts for uncertainty in using typical spectra in the calibration of the stochastic wind models.

% -----------------------------------------

\subsection{Model Errors}
\label{SimulationError}

To investigate model errors associated with the data-driven POD-based stochastic wind load model of Sec. \ref{Sect_DataDrivenPOD}, the simulation model was run with sample sizes ranging from 1,000 to 50,000 from which simulated spectra were estimated.  
%4-sec singles were generated and used to estimate the statistics of the process. The errors associated with the sample sizes are evaluated as described in Section \ref{error_measures}. 
The signals were generated using the stochastic wind model calibrated to the target spectra considering all 75 modes. Since the frequency intervals are identical to the input spectra, there is no interpolation involved thus ensuring the absence of interpolation errors.

\begin{figure}
	\centering
		\includegraphics[scale=.8]{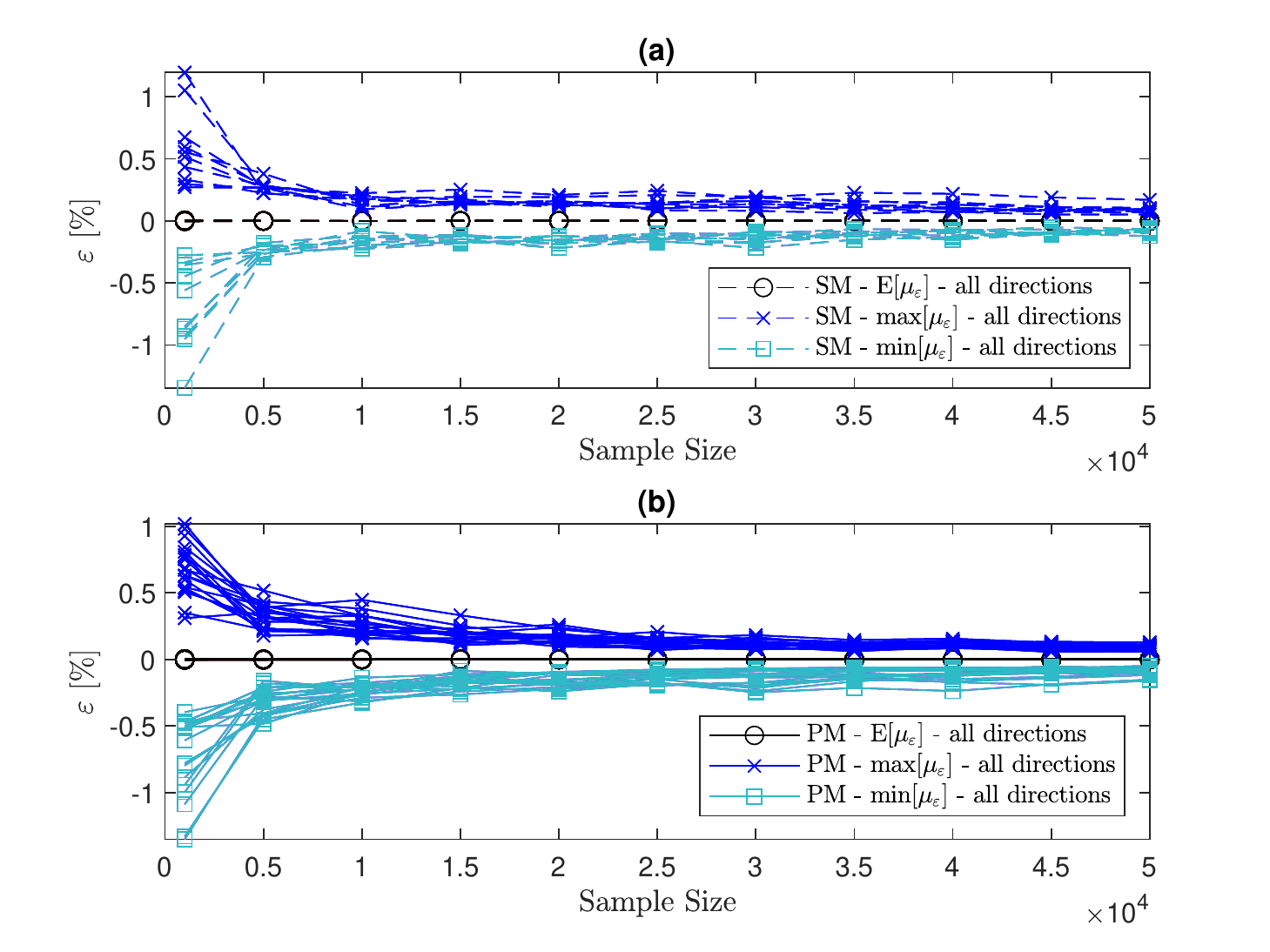}
	\caption{(a) The error in variance for the SM layout and various sample sizes, (b) the error in variance statistics for the PM layout and various sample sizes.}
	\label{varconvergence}
\end{figure}

Figures \ref{varconvergence} (a)-(b) report the trends in statistical errors in the variance when the sample size increases for both SM and PM layouts over all wind directions. A clear trend towards convergence can be observed. In particular, considering 50,000 samples as adequate for eliminating the majority of the statistical error (i.e., the error induced by sample size), a minimum and maximum mean model error, $\min[\mu_{\varepsilon}]$ and $\max[\mu_{\varepsilon}]$, of around -0.16\% and 0.17\% can be identified for both experimental layouts. 

\begin{figure}
	\centering
		\includegraphics[scale=.8]{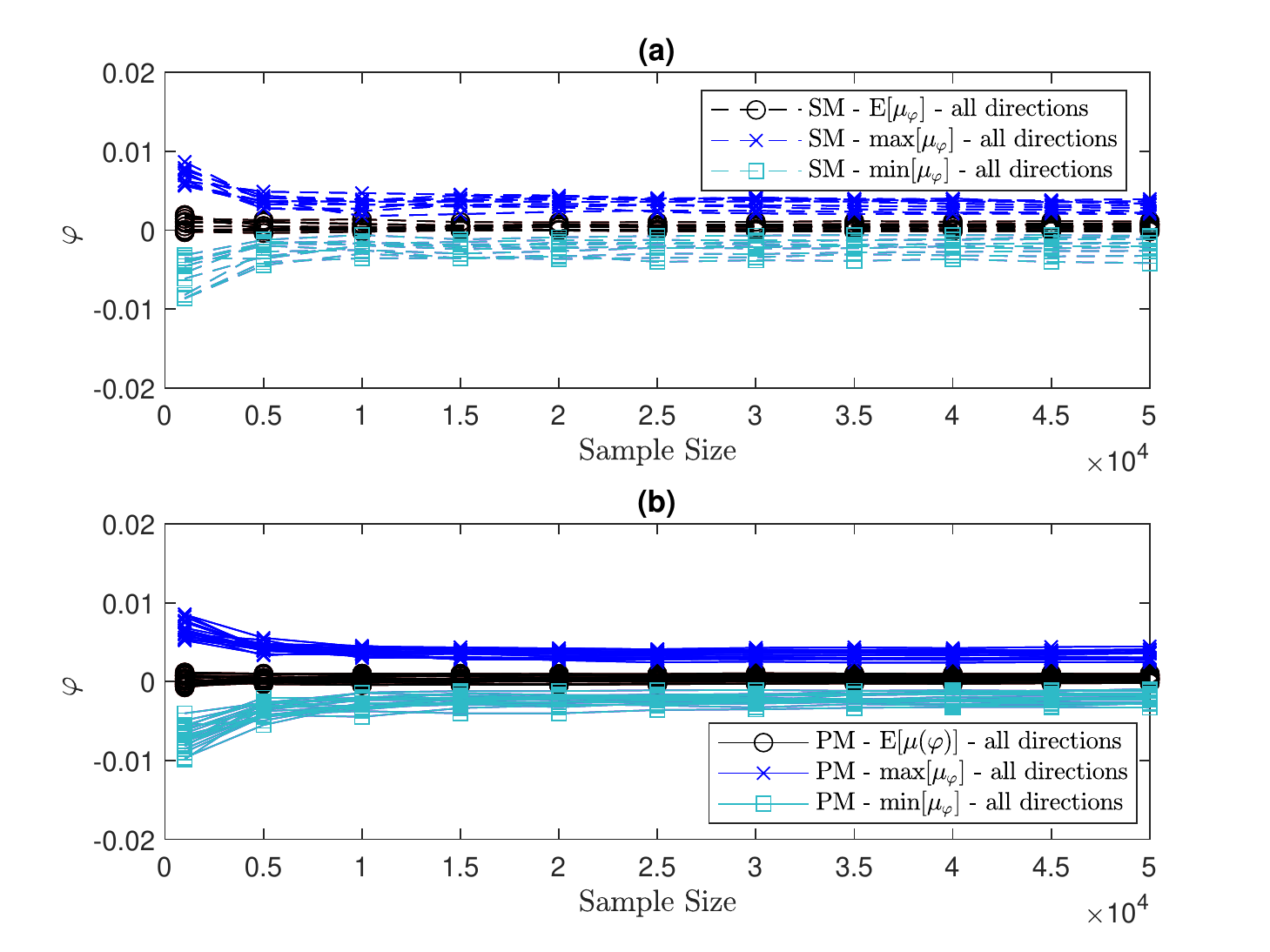}
	\caption{(a) The error in the correlation coefficient for the SM layout and various sample sizes, (b) the error in the correlation coefficients for the PM layout and various sample sizes.}
	\label{corrconvergence}
\end{figure}

Figure \ref{corrconvergence} (a)-(b) presents the model errors in the correlation coefficients for both layouts, with maximum and minimum mean error, $\min[\mu_{\varphi}]$ and $\max[\mu_{\varphi}]$, assuming values of around $-4\times 10^{-3}$ and $4.5\times 10^{-3}$ respectively. The expectation of errors are in the the order of $10^{-4}\%$ for $E[\mu_{\varepsilon}]$ and $10^{-4}$ for $E[\mu_{\varphi}]$.

\begin{figure}
	\centering
		\includegraphics[scale=.8]{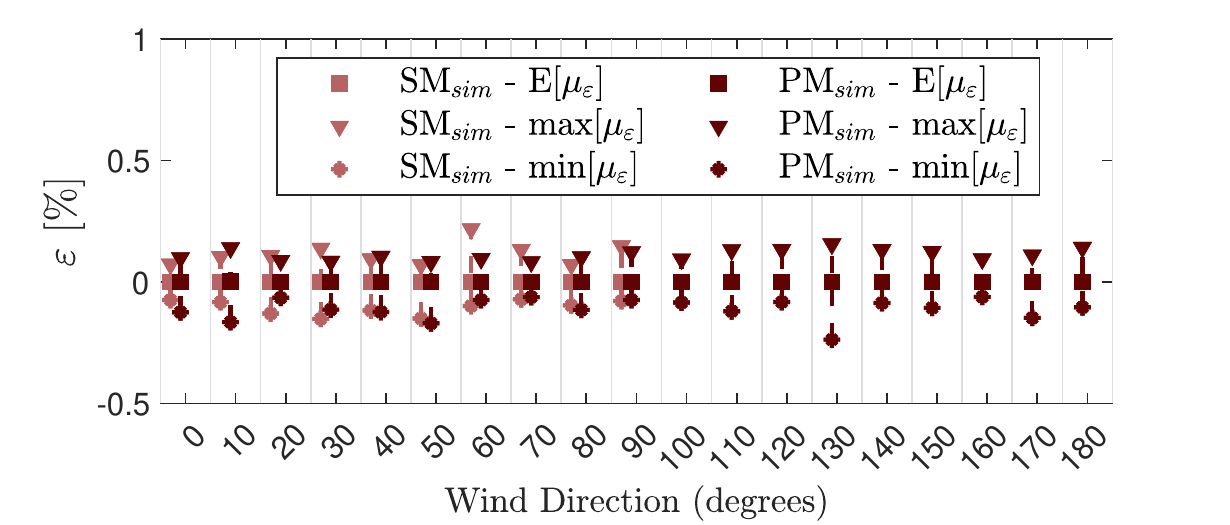}
	\caption{Expected values and range of the model error in variance for the SM and PM setup.}
	\label{sim_meanerror}
\end{figure}

To better illustrate the model errors in variance for individual wind directions, Fig. \ref{sim_meanerror} reports variations of the error measures for a sample size of 40,000, which was deemed sufficient to make any statistical errors negligible. It can be seen that the expected mean error, E$[\mu_{\varepsilon}]$, for all wind directions and settings is in the order of $10^{-4}$\%, with minimum and maximum mean errors equal to -0.23\% and 0.22\%, respectively. It is evident that the simulations follow the target spectra extremely well when all modes are considered. This result suggests that the data-driven POD-based stochastic wind load model operates with negligible model error irrespective of interference effects and wind direction.

\begin{figure}
	\centering
		\includegraphics[scale=.8, angle=90]{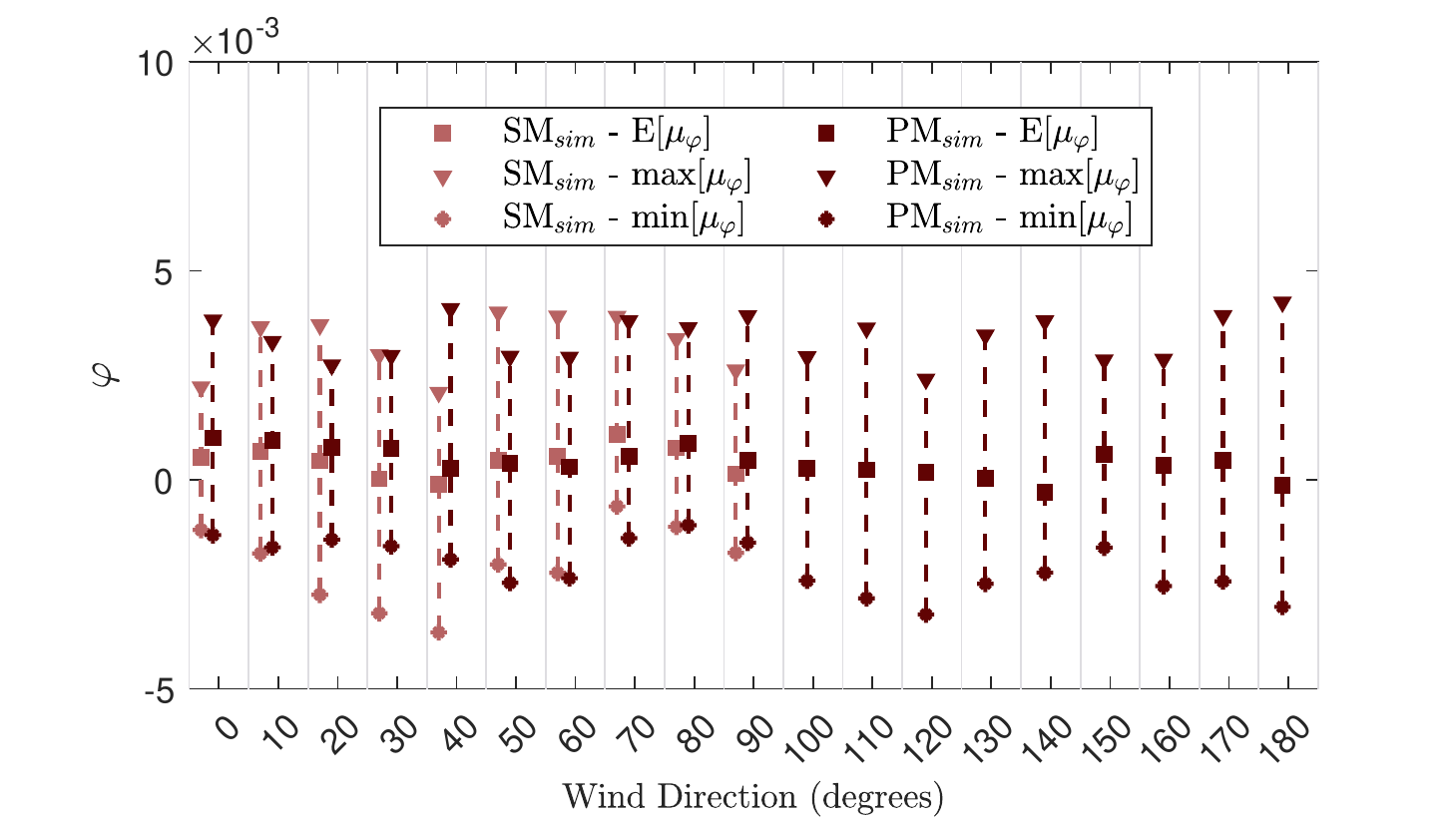}
	\caption{Expected values and range of the model error in correlation coefficients for the SM and PM setup.}
	\label{sim_diffall}
\end{figure}

Regarding the correlation coefficients, Fig. \ref{sim_diffall} shows the summary of the key statistics associated with $\mu_{\varphi}$. Similar to the error in variance, the value of E$[\mu_{\varphi}]$ also oscillates around 0, demonstrating that the correlation coefficients of the simulated signals agree well with the target correlation coefficients for all wind directions and experimental settings.

\begin{figure}
	\centering
		\includegraphics[scale=.8]{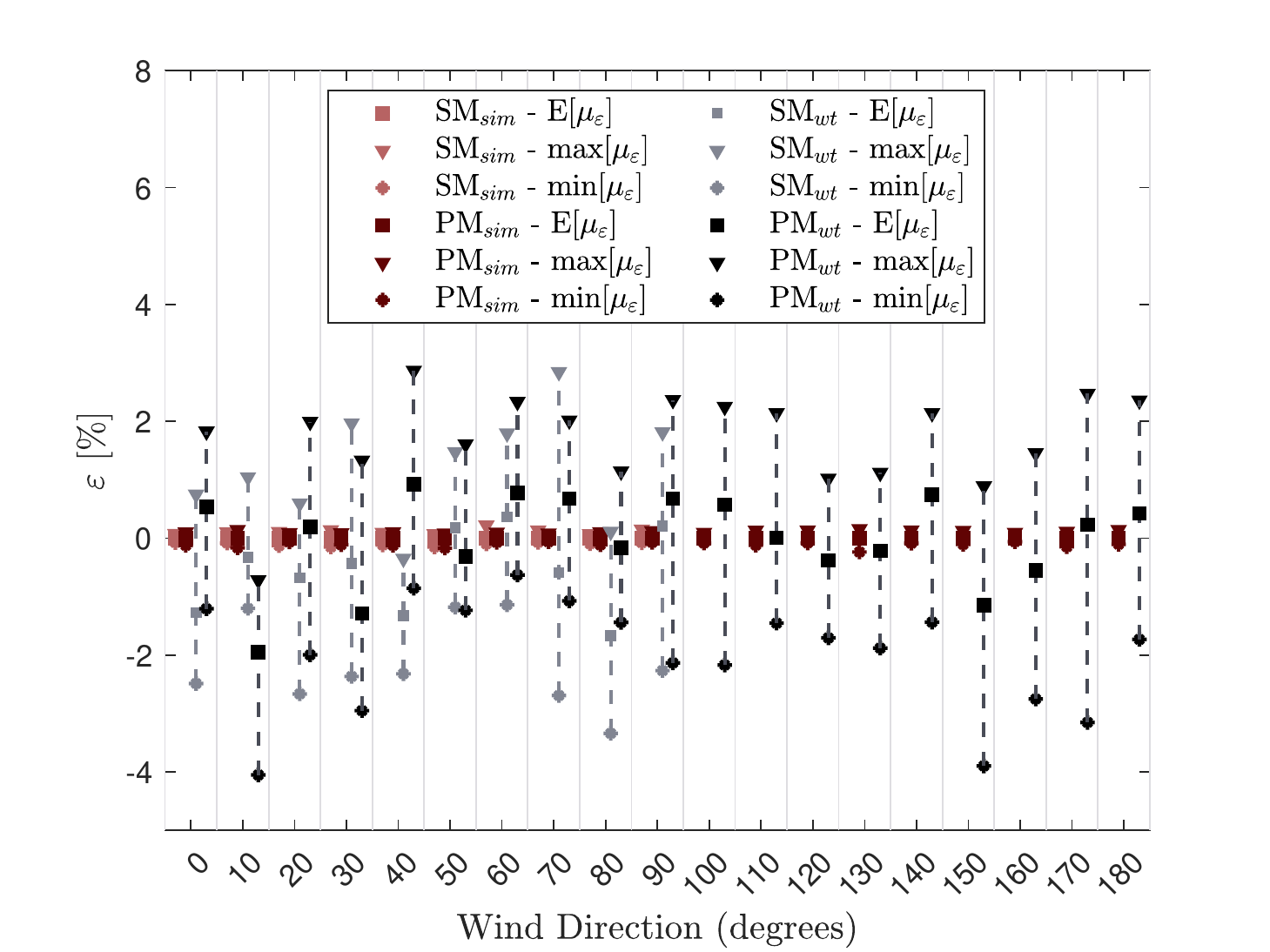}
	\caption{Comparison of the error in variance due to wind tunnel variability (wt) and the simulation model (sim).}
	\label{data_simulation}
\end{figure}

\begin{figure}
	\centering
		\includegraphics[scale=.8]{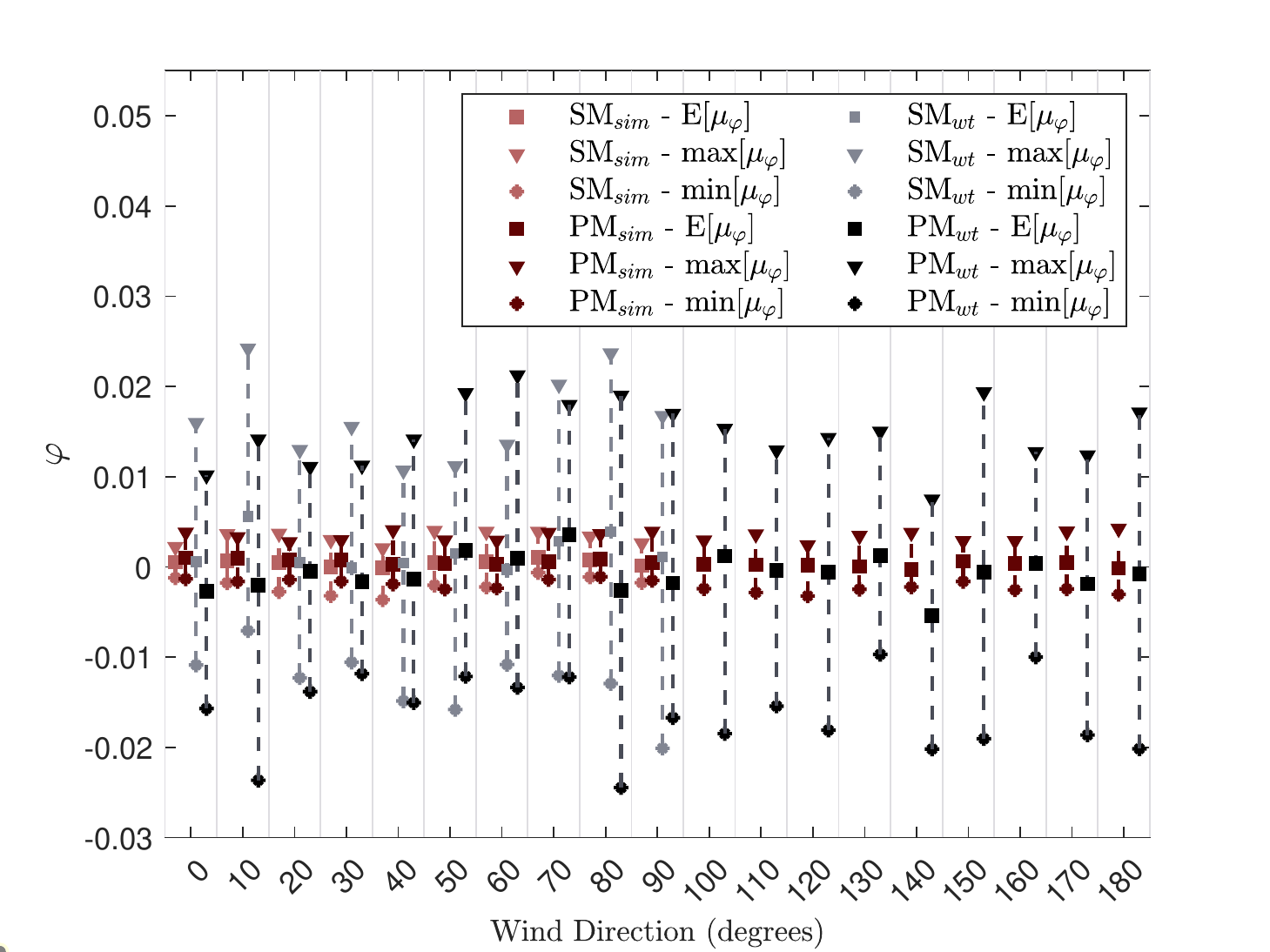}
	\caption{Comparison of the difference between correlation coefficients due to wind tunnel variability (wt) and the simulation model (sim).}
	\label{data_sim_diff}
\end{figure}

 Figures \ref{data_simulation} and \ref{data_sim_diff} examine the relative magnitude of the model errors as compared to those induced by the use of typical spectra. From these figures, it is evident that model errors are insignificant compared to the errors introduced by the use of typical wind tunnel records and therefore typical spectra during calibration. 

% -------------------------------------

\subsection{Truncation Error of POD-based simulation}
\label{TruncationError}

A parametric study on mode truncation was conducted to estimate the dependence of the errors in variance and correlation coefficients on the number of modes included in the simulation. Mode truncation can increase computational efficiency, hence analyzing truncation errors can help identify the appropriate number of modes to consider when adopting the data-driven POD-based stochastic wind load model. 
\begin{figure}
	\centering
		\includegraphics[scale=0.65]{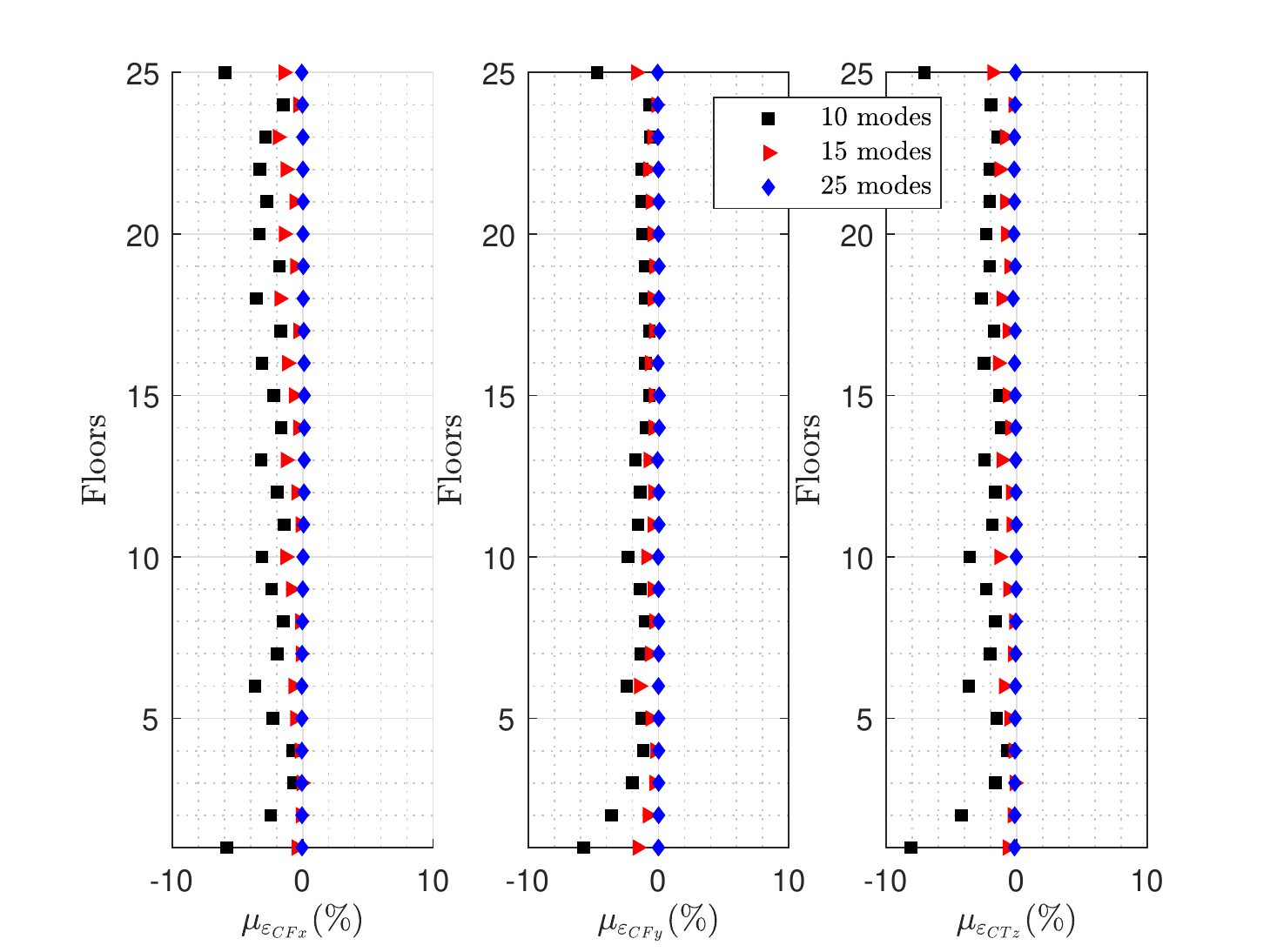}
	\caption{Mean error of variance, $\mu_{\varepsilon}$, for each force coefficient component considering 10, 15, and 25 contributing modes.}
	\label{10_20_modes_error}
\end{figure}
\begin{figure*}[bp]
		\includegraphics[scale=.8]{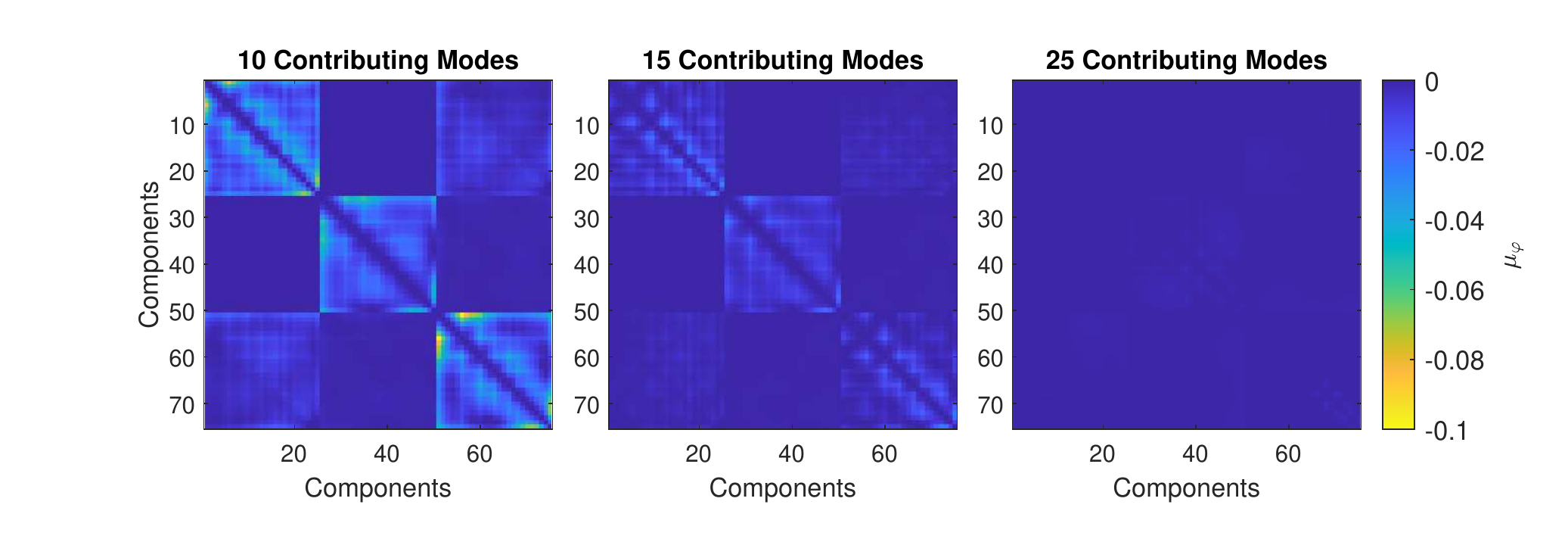}
	\caption{Difference between target and simulated signals considering 10, 15, and 25 modes.}
	\label{10_15_20modes_diff}
\end{figure*}

To illustrate the extent of the errors introduced by the truncation of modes, a total of 10, 15, and 25 out of the 75 total modes were considered, corresponding to 13\%, 20\%, and 33\% of all modes, respectively. The results for the SM layout and wind direction of $\beta=0^{\circ}$ are shown in Figs. \ref{10_20_modes_error}-\ref{10_15_20modes_diff}. The expected value of $\mu_{\varepsilon}$, shown in Fig. \ref{10_20_modes_error}, reduces as the number of modes increases, as expected. Figure \ref{10_15_20modes_diff} shows how the map of mean difference in correlation coefficients changes for an increase in the number of contributing modes. The results demonstrate that the consideration of the lower modes associated with higher magnitude eigenvalues is crucial to capture most of the energy, and the consideration of higher modes only improves the accuracy in increasing small amounts. 

\begin{figure}
	\centering
		\includegraphics[scale=.8, angle=90]{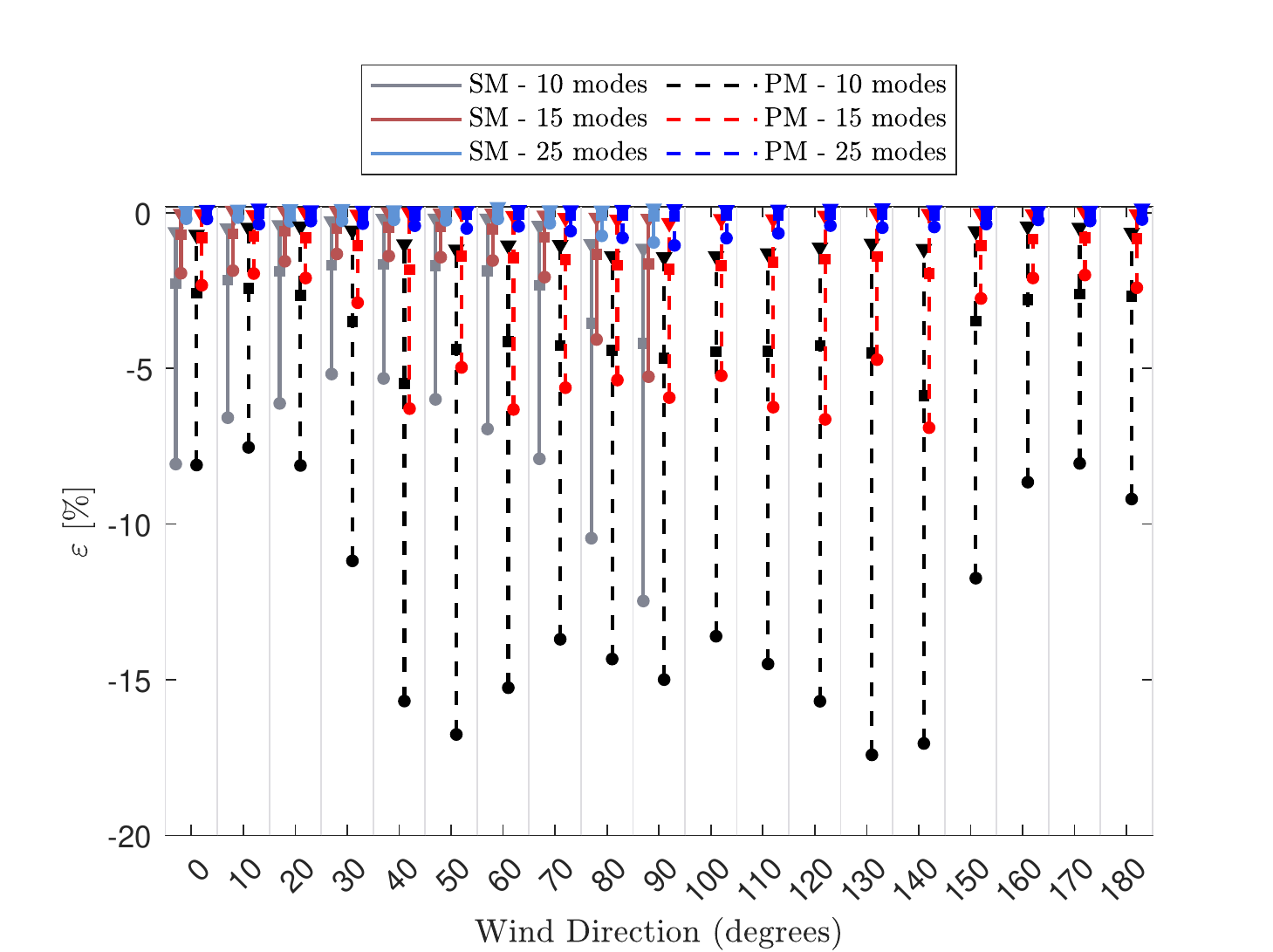}
	\caption{Error in variance, $\varepsilon$, considering 10, 15, and 25 contributing modes.}
	\label{meanerror_var_trunc}
\end{figure}

\begin{figure}
	\centering
		\includegraphics[scale=.8,angle=90]{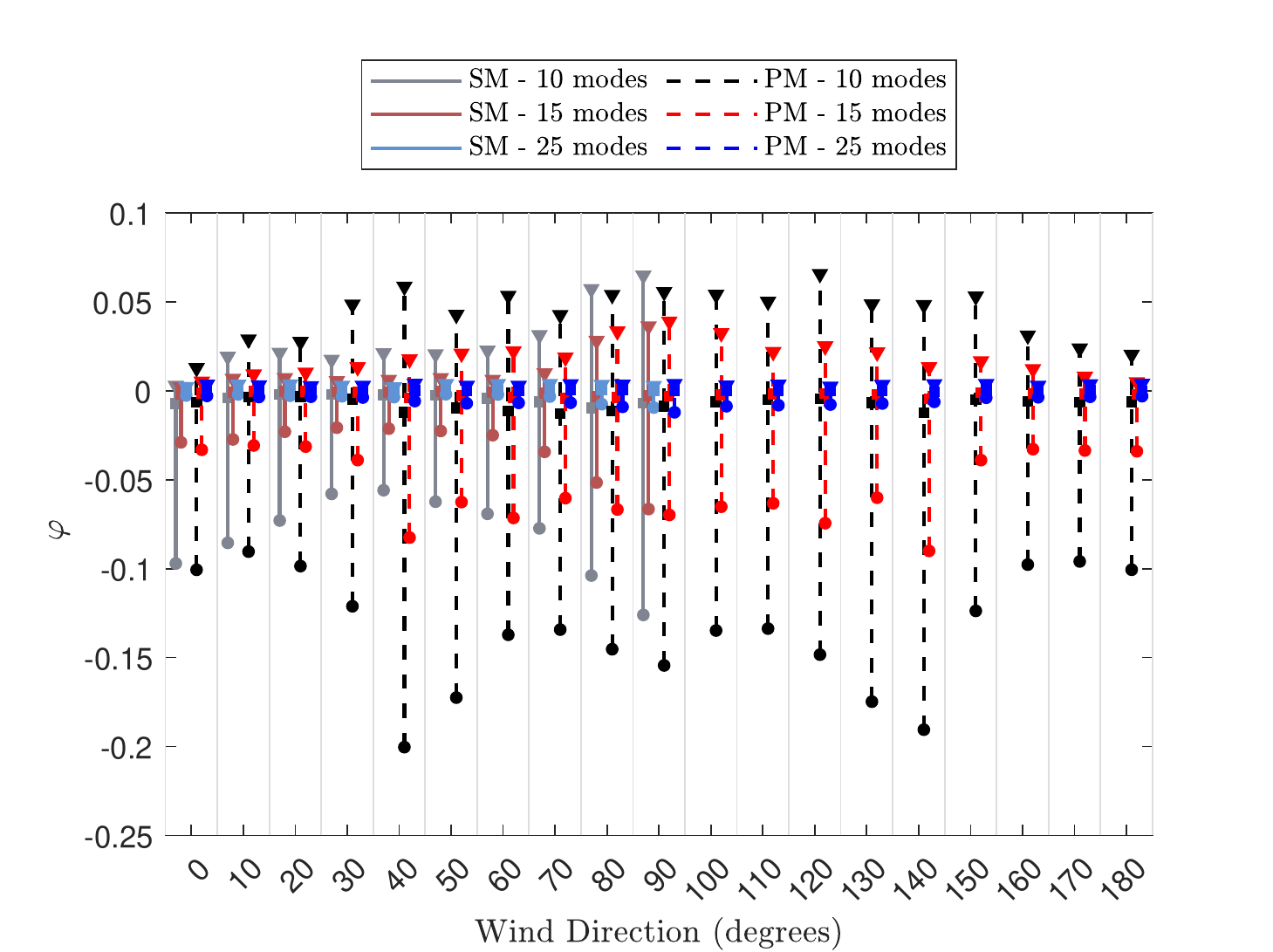}
	\caption{Difference between target and simulated signals, $\varphi$, considering 10, 15, and 25 contributing modes.}
	\label{meandif_corrcoef_trunc}
\end{figure}

Figures \ref{meanerror_var_trunc}-\ref{meandif_corrcoef_trunc} summarize the statistics of $\varepsilon$ and $\varphi$, respectively, for varying wind directions and configurations. For a reduced number of modes, i.e., 10 and 15 modes, the errors are considerably larger and more sensitive to changes in wind directions and configurations. However, for a sufficient number of contributing modes, i.e., 25 modes, the errors are sufficiently small for all cases. This suggests that a minimum number of modes in the simulation is necessary to ensure small and consistent errors given any wind direction and configuration. 

\begin{figure}
	\centering
		\includegraphics[scale=.8]{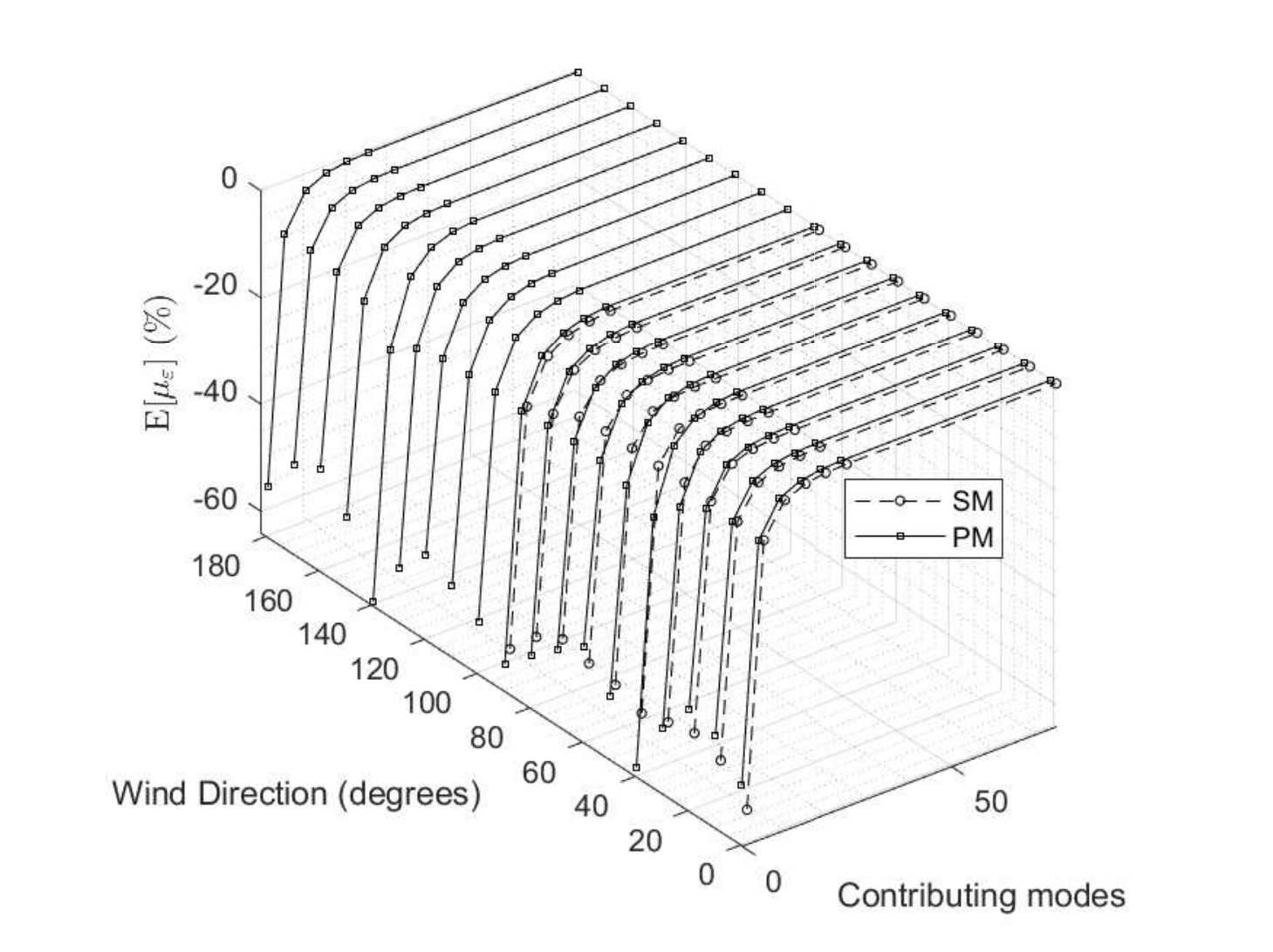}
	\caption{Expected error in variance from truncation of modes for both the SM and PM setups, considering 1, 5, 10, 15, 20, 25, and 75 contributing modes.}
	\label{modetruncation}
\end{figure}

\begin{figure}
	\centering
		\includegraphics[scale=.8]{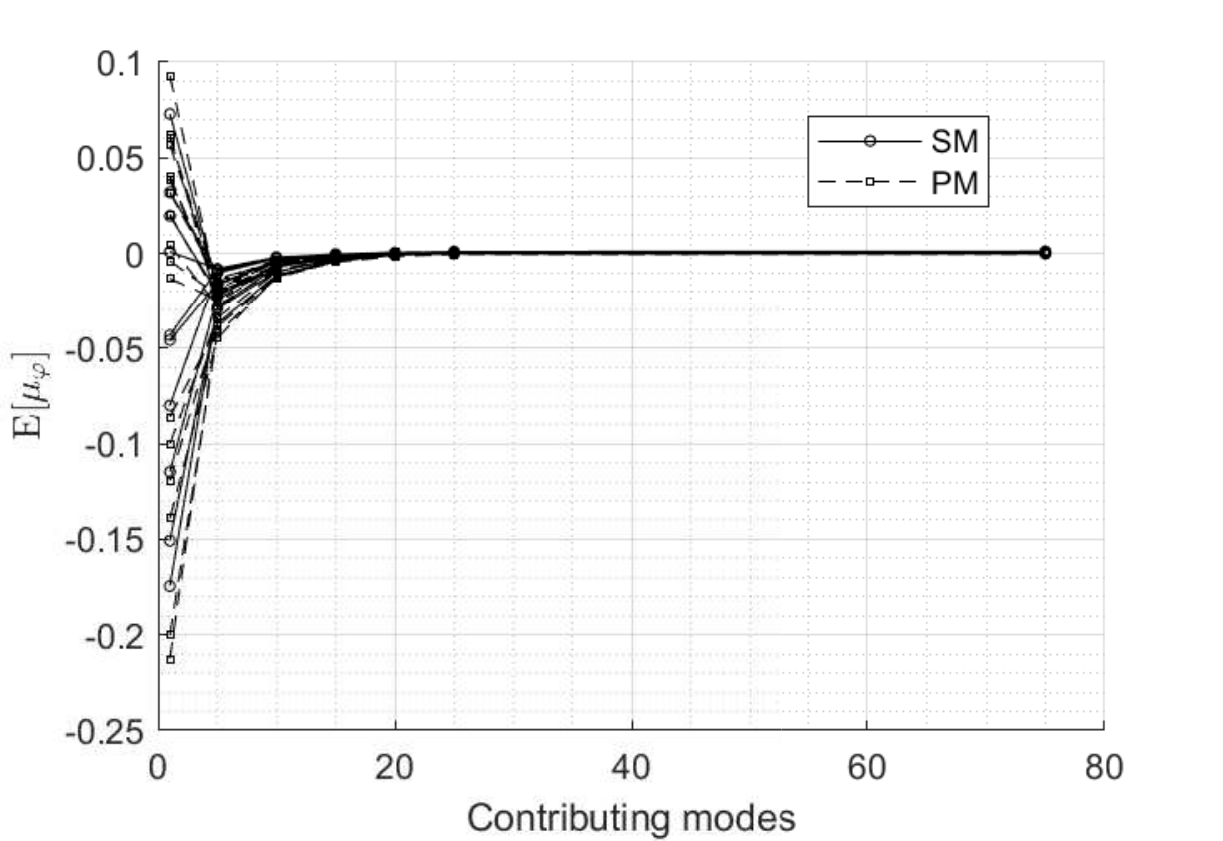}
	\caption{Expected difference in correlation coefficients from truncation of modes for all wind directions for both the SM and PM setups, considering 1, 5, 10, 15, 20, 25, and 75 contributing modes.}
	\label{modetruncation_corr}
\end{figure}

Figures \ref{modetruncation} and \ref{modetruncation_corr} show the expected values of $\mu_{\varepsilon}$ and the expected values of $\mu_{\varphi}$ for all wind directions and both experimental settings for varying numbers of contributing modes. Results demonstrate that including a higher number of modes clearly increases the accuracy of simulated signals. E$[\mu_{\varepsilon}]$ reaches a value between -0.2\% to -0.6\% for approximately 20 modes (i.e., around 27\% of all modes) and E$[\mu_{\varphi}]$ reaches a value of around $10^{-4}$, which is comparable to the model error considering all modes. This means that higher modes do not have a significant contribution to the loading process and can be truncated with no noticeable loss of accuracy. 

% ---------------------------------
% ---------------------------------
% ---------------------------------

\section{Conclusions and Discussion}

An extensive wind tunnel test campaign was conducted considering multiple wind directions and two different experimental configurations to ultimately investigate the applicability of a wind tunnel-informed POD-based stochastic wind model and quantify the errors associated with calibration to typical wind tunnel data, the model itself, and mode truncation. The data were divided into two groups, one group was used for defining the target spectra of the wind load process, while the other was used for estimating the spectra from typical wind tunnel records. 

The results of this work demonstrated that data-driven stochastic wind load models can efficiently simulate stochastic wind loads with negligible model error, while the errors induced from calibration to limited wind tunnel data can be considerable but can be reduced through the use of adequate experimental data. Regarding mode truncation, it was seen that approximately 25\% of modes should be included to guarantee a negligible loss of accuracy. It is hoped that the extensive error analysis associated with the use of the data-driven POD-based stochastic wind load model of this work will provide greater confidence for the use of such models in wind engineering applications.

% ---------------------------------
% ---------------------------------
% ---------------------------------

\appendix

\section{Standardization scheme of force coefficients}

To increase the efficiency of carrying out a POD analysis of wind force coefficients, it is first advisable to standardize the data. To this end, the zero-mean force coefficients were standardized in this study using the following scheme:
\begin{equation}
   CF_{x,n}^{norm}(t)=\frac{CF_{x,n}(t)}{\sigma_{CF_{x,n}}\gamma_r}
\end{equation}
\begin{equation}
   CF_{y,n}^{norm}(t)=\frac{CF_{y,n}(t)}{\sigma_{CF_{y,n}}\gamma_r}
\end{equation}
\begin{equation}
   CT_{z,n}^{norm}(t)=\frac{CT_{z,n}(t)}{\sigma_{CT_{z,n}}\gamma_r}
\end{equation}
where $CF_x(t)$, $CF_y(t)$, $CT_z(t)$, are the zero-mean force coefficients while $\gamma_r=3.5$ is the reduced variate estimated based on the expected peak of a Gaussian process after normalization. 

%After the simulation, it is straightforward to :
%
%\begin{equation}
%   CF_{x,n}^{sim}(t)=CF_{x,n}^{s,norm}(t)\sigma_{CF_{x,n}}\gamma_r
%\end{equation}
%\begin{equation}
%   CF_{y,n}^{sim}(t)=CF_{y,n}^{s,norm}(t)\sigma_{CF_{y,n}}\gamma_r
%\end{equation}
%\begin{equation}
%   CT_{z,n}^{sim}(t)=CT_{z,n}^{s,norm}(t)\sigma_{CT_{z,n}}\gamma_r
%\end{equation}
%
%where $CF_{x,n}^{s,norm}(t)$, $CF_{y,n}^{s,norm}(t)$, $CT_{z,n}^{s,norm}(t)$ are the simulated signals using the wind-tunnel informed POD-based stochastic wind load model.

\section*{Declaration of competing interest}
The authors declare that they have no known competing financial interests or personal relationships that could have appeared to influence the work reported in this paper.

\section*{Acknowledgements}
The authors would like to gratefully acknowledge the support of the Natural Science Foundation (Grants: CMMI-1750339, CMMI-2118488, and CMMI-2131111), and acknowledge Dr. Sungmoon Jung for providing the scale model used in this study.

\section*{Data Availability Statements}
The wind tunnel data generated for this study is published in the Design Safe Database and can be accessed free of charge.

%% Loading bibliography style file
\newpage
\bibliographystyle{unsrt}
%\bibliographystyle{cas-model2-names}

% Loading bibliography database
\bibliography{ms}

\begin{thebibliography}{10}

\bibitem{ciampoli2011performance}
M~Ciampoli, F~Petrini, and G~Augusti.
\newblock Performance-based wind engineering: towards a general procedure.
\newblock {\em Structural Safety}, 33(6):367--378, 2011.

\bibitem{smith2011monte}
Marra~A Smith and Luca Caracoglia.
\newblock A monte carlo based method for the dynamic “fragility analysis”
  of tall buildings under turbulent wind loading.
\newblock {\em Engineering Structures}, 33(2):410--420, 2011.

\bibitem{petrini2012performance}
Francesco Petrini and Marcello Ciampoli.
\newblock Performance-based wind design of tall buildings.
\newblock {\em Structure and Infrastructure Engineering}, 8(10):954--966, 2012.

\bibitem{barbato2013performance}
Michele Barbato, Francesco Petrini, Vipin~U Unnikrishnan, and Marcello
  Ciampoli.
\newblock Performance-based hurricane engineering (pbhe) framework.
\newblock {\em Structural Safety}, 45:24--35, 2013.

\bibitem{spence2014performance}
Seymour M~J Spence and Ahsan Kareem.
\newblock Performance-based design and optimization of uncertain wind-excited
  dynamic building systems.
\newblock {\em Engineering Structures}, 78:133--144, 2014.

\bibitem{bernardini2015performance}
Enrica Bernardini, Seymour M~J Spence, Dae-Kun Kwon, and Ahsan Kareem.
\newblock Performance-based design of high-rise buildings for occupant comfort.
\newblock {\em Journal of Structural Engineering}, 141(10):04014244, 2015.

\bibitem{chuang2017performance}
Wei-Chu Chuang and Seymour M~J Spence.
\newblock A performance-based design framework for the integrated collapse and
  non-collapse assessment of wind excited buildings.
\newblock {\em Engineering Structures}, 150:746--758, 2017.

\bibitem{cui2018unified}
Wei Cui and Luca Caracoglia.
\newblock A unified framework for performance-based wind engineering of tall
  buildings in hurricane-prone regions based on lifetime intervention-cost
  estimation.
\newblock {\em Structural Safety}, 73:75--86, 2018.

\bibitem{ouyang2020performance}
Zhicheng Ouyang and Seymour M~J Spence.
\newblock A performance-based wind engineering framework for envelope systems
  of engineered buildings subject to directional wind and rain hazards.
\newblock {\em Journal of Structural Engineering}, 146(5):04020049, 2020.

\bibitem{Cui_2020}
W.~Cui and L.~Caracoglia.
\newblock Performance-based wind engineering of tall buildings examining
  life-cycle downtime and multisource wind damage.
\newblock {\em J. Struct. Eng.}, 146(1), 2020.

\bibitem{Ouyang_21}
Z.~Ouyang and S.~M.~J. Spence.
\newblock Performance-based wind-induced structural and envelope damage
  assessment of engineered buildings through nonlinear dynamic analysis.
\newblock {\em Journal of Wind Engineering and Industrial Aerodynamics},
  208(1):104452, 2021.

\bibitem{Arunachalam_2022}
S.~Arunachalam and S.~M.~J. Spence.
\newblock Reliability-based collapse assessment of wind-excited steel
  structures within performance-based wind engineering.
\newblock {\em Journal of Structural Engineering}, 148(9):04022132, 2022.

\bibitem{chuang2022framework}
Wei-Chu Chuang and Seymour M.~J. Spence.
\newblock A framework for the efficient reliability assessment of inelastic
  wind excited structures at dynamic shakedown.
\newblock {\em Journal of Wind Engineering and Industrial Aerodynamics},
  220:104834, 2022.

\bibitem{lin2005characteristics}
Ning Lin, Chris Letchford, Yukio Tamura, Bo~Liang, and Osamu Nakamura.
\newblock Characteristics of wind forces acting on tall buildings.
\newblock {\em Journal of Wind Engineering and Industrial Aerodynamics},
  93(3):217--242, 2005.

\bibitem{kareem2013advanced}
Yukio Tamura and Ahsan Kareem.
\newblock {\em Advanced Structural Wind Engineering}, volume 482.
\newblock Springer, 2013.

\bibitem{shinozuka1971simulation}
Masanobu Shinozuka.
\newblock Simulation of multivariate and multidimensional random processes.
\newblock {\em The Journal of the Acoustical Society of America},
  49(1B):357--368, 1971.

\bibitem{shinozuka1991simulation}
Masanobu Shinozuka and George Deodatis.
\newblock Simulation of stochastic processes by spectral representation.
\newblock {\em Applied Mechanics Reviews}, 44(4):191--204, 1991.

\bibitem{deodatis1996simulation}
George Deodatis.
\newblock Simulation of ergodic multivariate stochastic processes.
\newblock {\em Journal of Engineering Mechanics}, 122(8):778--787, 1996.

\bibitem{tamura1999proper}
Y~Tamura, S~Suganuma, H~Kikuchi, and K~Hibi.
\newblock Proper orthogonal decomposition of random wind pressure field.
\newblock {\em Journal of Fluids and Structures}, 13(7-8):1069--1095, 1999.

\bibitem{carassale2001double}
Luigi Carassale, Giuseppe Piccardo, and Giovanni Solari.
\newblock Double modal transformation and wind engineering applications.
\newblock {\em Journal of Engineering Mechanics}, 127(5):432--439, 2001.

\bibitem{chen2005simulation}
Lizhong Chen and Chris~W Letchford.
\newblock Simulation of multivariate stationary gaussian stochastic processes:
  Hybrid spectral representation and proper orthogonal decomposition approach.
\newblock {\em Journal of Engineering Mechanics}, 131(8):801--808, 2005.

\bibitem{chen2005proper}
Xinzhong Chen and Ahsan Kareem.
\newblock Proper orthogonal decomposition-based modeling, analysis, and
  simulation of dynamic wind load effects on structures.
\newblock {\em Journal of Engineering Mechanics}, 131(4):325--339, 2005.

\bibitem{huang2020data}
Guoqing Huang, Liuliu Peng, Ahsan Kareem, and Chunchen Song.
\newblock Data-driven simulation of multivariate nonstationary winds: A hybrid
  multivariate empirical mode decomposition and spectral representation method.
\newblock {\em Journal of Wind Engineering and Industrial Aerodynamics},
  197:104073, 2020.

\bibitem{tao2020error}
Tianyou Tao, Hao Wang, Liang Hu, and Ahsan Kareem.
\newblock Error analysis of multivariate wind field simulated by
  interpolation-enhanced spectral representation method.
\newblock {\em Journal of Engineering Mechanics}, 146(6):04020049, 2020.

\bibitem{hu2010error}
Liang Hu, Li~Li, and Ming Gu.
\newblock Error assessment for spectral representation method in wind velocity
  field simulation.
\newblock {\em Journal of Engineering Mechanics}, 136(9):1090--1104, 2010.

\bibitem{tao2018reduced}
Tianyou Tao, Hao Wang, and Ahsan Kareem.
\newblock Reduced-hermite bifold-interpolation assisted schemes for the
  simulation of random wind field.
\newblock {\em Probabilistic Engineering Mechanics}, 53:126--142, 2018.

\bibitem{davenport1971response}
AG~Davenport.
\newblock The response of six building shapes to turbulent wind.
\newblock {\em Philosophical Transactions of the Royal Society of London A,
  Mathematical, Physical and Engineering Sciences}, 269(1199):385--394, 1971.

\bibitem{simiu1974wind}
Emil Simiu.
\newblock Wind spectra and dynamic alongwind response.
\newblock {\em Journal of the Structural Division}, 100(9):1897--1910, 1974.

\bibitem{melbourne1980comparison}
WH~Melbourne.
\newblock Comparison of measurements on the caarc standard tall building model
  in simulated model wind flows.
\newblock {\em Journal of Wind Engineering and Industrial Aerodynamics},
  6(1-2):73--88, 1980.

\bibitem{solari1983analytical}
G~Solari.
\newblock Analytical estimation of the alongwind response of structures.
\newblock {\em Journal of Wind Engineering and Industrial Aerodynamics},
  14(1-3):467--477, 1983.

\bibitem{kaimal1994atmospheric}
Jagadish~Chandran Kaimal and John~J Finnigan.
\newblock {\em Atmospheric boundary layer flows: their structure and
  measurement}.
\newblock Oxford University Press, 1994.

\bibitem{kareem2008numerical}
Ahsan Kareem.
\newblock Numerical simulation of wind effects: A probabilistic perspective.
\newblock {\em Journal of Wind Engineering and Industrial Aerodynamics},
  96(10-11):1472--1497, 2008.

\bibitem{gurley1998simulation}
K~Gurley and A~Kareem.
\newblock Simulation of correlated non-gaussian pressure fields.
\newblock {\em Meccanica}, 33(3):309--317, 1998.

\bibitem{wang2013data}
Lijuan Wang, Megan McCullough, and Ahsan Kareem.
\newblock A data-driven approach for simulation of full-scale downburst wind
  speeds.
\newblock {\em Journal of Wind Engineering and Industrial Aerodynamics},
  123:171--190, 2013.

\bibitem{suksuwan2018optimization}
Arthriya Suksuwan and Seymour M~J Spence.
\newblock Optimization of uncertain structures subject to stochastic wind loads
  under system-level first excursion constraints: a data-driven approach.
\newblock {\em Computers \& Structures}, 210:58--68, 2018.

\bibitem{american2019prestandard}
ASCE.
\newblock Prestandard for performance-based wind design.
\newblock American Society of Civil Engineers, 2019.

\bibitem{gurley1997modelling}
Kurtis~Robert Gurley.
\newblock {\em Modelling and simulation of non-Gaussian processes}.
\newblock University of Notre Dame, 1997.

\bibitem{shinozuka1987stochastic}
M~Shinozuka.
\newblock Stochastic fields and their digital simulation.
\newblock In {\em Stochastic Methods in Structural Dynamics}, pages 93--133.
  Springer, 1987.

\bibitem{solomon1991psd}
Otis~M Solomon~Jr.
\newblock Psd computations using welch's method.
\newblock Technical report, Sandia National Labs., Albuquerque, NM (United
  States), 1991.

\bibitem{Welch}
P.~Welch.
\newblock The use of fast fourier transform for the estimation of power
  spectra: A method based on time averaging over short, modified periodograms.
\newblock {\em IEEE Transactions on Audio and Electroacoustics}, 15(2):70--73,
  1967.

\bibitem{huppenkothen2018statistical}
D~Huppenkothen and Matteo Bachetti.
\newblock On the statistical properties of cospectra.
\newblock {\em The Astrophysical Journal Supplement Series}, 236(1):13, 2018.

\bibitem{catarelli2020automation}
Ryan~A Catarelli, Pedro~L Fern{\'a}ndez-Cab{\'a}n, Brian~M Phillips, Jennifer~A
  Bridge, Forrest~J Masters, Kurtis~R Gurley, and David~O Prevatt.
\newblock Automation and new capabilities in the university of florida nheri
  boundary layer wind tunnel.
\newblock {\em Frontiers in Built Environment}, 6, 2020.

\end{thebibliography}

\end{document}